%% file: main.tex
\documentclass[a4paper,11pt]{article}
\usepackage{jinstpub} 
\usepackage{lineno}
\usepackage{jinstpub} 
\usepackage{comment}

\usepackage[utf8]{inputenc}
\usepackage[figuresright]{rotating}
\usepackage[section]{placeins}
\usepackage{bm} 
\usepackage{color}
\usepackage{graphicx}
\usepackage{hyperref} 
\usepackage{latexsym}
\usepackage{relsize}
\usepackage{soul}
\usepackage{subcaption}
\usepackage{xspace}
\usepackage[colorinlistoftodos]{todonotes}
\usepackage{placeins}
\usepackage[version=4]{mhchem}
\usepackage{graphicx} 
\usepackage{xcolor}
\usepackage{gensymb}
\usepackage{array}
\newcounter{daggerfootnote}
\newcommand*{\daggerfootnote}[1]{%
    \setcounter{daggerfootnote}{\value{footnote}}%
    \renewcommand*{\thefootnote}{\fnsymbol{footnote}}%
    \footnote[2]{#1}%
    \setcounter{footnote}{\value{daggerfootnote}}%
    \renewcommand*{\thefootnote}{\arabic{footnote}}%
    }

\keywords{Vacuum-based detectors, Detector modelling and simulations II (electric fields, charge transport, multiplication and induction, pulse formation, electron emission, etc), Trigger detectors, Instrumentation and methods for time-of-flight (TOF) spectroscopy}

\arxivnumber{2503.10025} 
\title{\boldmath A Demonstration of Slowed Electron ${\bf E}$ $\times$ ${\bf B}$ Drift for PTOLEMY}

\author[1]{M.~Farino,\daggerfootnote{Corresponding author.}}
\author[1]{A.~Tan,}
\author{and the PTOLEMY Collaboration:}
\author[2,3]{A.~Apponi,}
\author[4,5]{M.~Betti,}
\author[6,7]{M.~Borghesi,}
\author[8]{A.~Casale,}
\author[4,5]{O.~Castellano,}
\author[4,5]{G.~Cavoto,}
\author[4]{L.~Cecchini,}
\author[9]{E.~Celasco,}
\author[1]{W.~Chung,}
\author[10]{A.G.~Cocco,}
\author[11,12]{A.~Colijn,}
\author[4,5]{B.~Corcione,}
\author[10]{N.~D'Ambrosio,}
\author[13]{N.~de~Groot,}
\author[12]{S.~el~Morabit,}
\author[4,5]{A.~Esposito,}
\author[6,7]{M.~Faverzani,}
\author[10,14]{A.D.~Ferella,}
\author[6]{E.~Ferri,}
\author[4,5]{L.~Ficcadenti,}
\author[6,7]{S.~Gamba,}
\author[15]{S.~Gariazzo,}
\author[16,17,18]{H.~Garrone,}
\author[9]{F.~Gatti,}
\author[6,7]{A.~Giachero,}
\author[19]{Y.~Iwasaki,}
\author[20]{A.~Kievsky,}
\author[17,18,21]{F.~Malnati}
\author[22,23]{G.~Mangano,}
\author[20,24]{ L.E.~Marcucci,}
\author[4,5]{C.~Mariani,}
\author[12]{J.~Mead,}
\author[24,25]{G.~Menichetti,}
\author[10]{M.~Messina,}
\author[17,18]{E.~Monticone,}
\author[11]{M.~Naafs,}
\author[6,7]{A.~Nucciotti,}
\author[6,7]{L.~Origo,}
\author[4]{F.~Pandolfi,}
\author[2,3]{D.~Paoloni,}
\author[17,26]{C.~Pepe,}
\author[27]{C.~Pérez~de~los~Heros,}
\author[22,23]{O.~Pisanti,}
\author[10, 28]{F.M.~Pofi,}
\author[4,5]{A.D.~Polosa,}
\author[4]{I.~Rago,}
\author[17,18]{M.~Rajteri,}
\author[10]{N.~Rossi,}
\author[2,3]{A.~Ruocco,}
\author[4,5]{S.~Tayyab,}
\author[20,29]{V.~Tozzini,}
\author[1]{C.~Tully,}
\author[13]{I.~van~Rens,}
\author[10,14]{F.~Virzi,}
\author[11]{G.~Visser,}
\author[20]{M.~Viviani,}
\author[13]{U.~Zeitler,}
\author[13]{O.~Zheliuk,}
\author[12]{F.~Zimmer}

\affiliation[1]{Princeton University, Princeton NJ, USA}
\affiliation[2]{Istituto Nazionale di Fisica Nucleare (INFN), Sezione di Roma Tre, Roma, Italy}
\affiliation[3]{Dipartimento di Scienze, Universit\`a degli Studi di Roma Tre, Roma, Italy}
\affiliation[4]{Istituto Nazionale di Fisica Nucleare (INFN), Sezione di Roma, Roma, Italy}
\affiliation[5]{Sapienza Universit\`a di Roma, Roma, Italy}
\affiliation[6]{Istituto Nazionale di Fisica Nucleare (INFN), Sezione di Milano Bicocca, Milano, Italy}
\affiliation[7]{Dipartimento di Fisica, Universit\`a di Milano, Bicocca, Milano, Italy}
\affiliation[8]{Department of Physics, Columbia University, New York NY, USA}
\affiliation[9]{Universit\`a di Genova e INFN Sezione di Genova, Genova, Italy}
\affiliation[10]{Istituto Nazionale di Fisica Nucleare (INFN) Laboratori Nazionali del Gran Sasso, L'Aquila, Italy}
\affiliation[11]{Nationaal instituut voor subatomaire fysica (Nikhef), Amsterdam, The Netherlands}
\affiliation[12]{University of Amsterdam, Amsterdam, The Netherlands}
\affiliation[13]{Radbound University, Nijmegen, The Netherlands }
\affiliation[14]{Universit\`a degli Studi dell'Aquila, L'Aquila, Italy}
\affiliation[15]{Instituto de F\'isica Te\'orica UAM-CSIC, Madrid, Spain}
\affiliation[16]{Department of Electronics and Telecommunications, Politecnico di Torino, Torino, Italy}
\affiliation[17]{Istituto Nazionale di Ricerca Metrologica (INRiM), Torino, Italy}
\affiliation[18]{Istituto Nazionale di Fisica Nucleare (INFN), Sezione di Torino, Torino, Italy}
\affiliation[19]{University of California, Berkeley CA, USA}
\affiliation[20]{Istituto Nazionale di Fisica Nucleare (INFN), Sezione di Pisa, Pisa, Italy}
\affiliation[21]{Department of Applied Science and Technology, Politecnico di Torino, Torino, Italy}
\affiliation[22]{Istituto Nazionale di Fisica Nucleare (INFN), Sezione di Napoli, Napoli, Italy}
\affiliation[23]{Universit\`a degli Studi di Napoli Federico II, Napoli, Italy}
\affiliation[24]{Department of Physics, University of Pisa, Pisa, Italy}
\affiliation[25]{Istituto Italiano di Tecnologia, Graphene Labs, Genova, Italy}
\affiliation[26]{Institut de Microelectronica de Barcelona, IMB-CNM (CSIC), Barcelona, Spain}
\affiliation[27]{Uppsala University, Uppsala, Sweden}
\affiliation[28]{Gran Sasso Science Institute (GSSI),  L'Aquila, Italy}
\affiliation[29]{Istituto Nanoscienze-CNR, Lab NEST-SNS, Pisa, Italy}

\emailAdd{mfarino@princeton.edu}

\abstract{{\color{black}To resolve the effective neutrino mass $m_\beta$ with an energy resolution of {\color{black} 50~meV}, the PTOLEMY experiment has proposed a novel transverse electromagnetic filtering process. {\color{black}Substantially} reducing the kinetic energy of tritium $\beta$-decay electrons by counteracting motion from ${\bf E}$~$\times$~${\bf B}$ and $\nabla{\rm B}$ drift, the PTOLEMY filter requires an input of emitted electron kinematic information to generate a tailored, suitable electric field for each candidate. The collaboration proposes to extract these quantities by using antennae to observe the relativistic frequency shift of emitted cyclotron radiation as an electron transits by ${\bf E}$ $\times$ ${\bf B}$ drift through a uniform magnetic field region preceding the filter. Electrons must be contained within this region long enough such that an adequate integrated radiated power signal is received to accurately estimate {\color{black} these} kinematics. This necessitates a controlled, slowed drift speed.} This paper presents the experimental design to vary ${\bf E}$ $\times$ ${\bf B}$ drift speed of $\ce{^{14}C}$ $\beta$-decay electrons using a custom electrode field cage situated between the pole faces of an electromagnet. Matching our results with high-fidelity simulation, we deduce a capacity to increase particle time of flight by a factor of 5 in the field cage's slow drift region. Limited only by the dimensions of our system, we assert drift speed can be arbitrarily slowed to meet the needs of PTOLEMY's future detector. {\color{black}Actualizing such a system is a crucial milestone in developing the detector, enabling future cyclotron radiation measurements, filter implementation, and source injection.}}

\begin{document}

\maketitle

\flushbottom

\section{Introduction}

\subsection {Neutrino Mass Landscape}
{\color{black}Juxtaposed to the Standard Model of particle physics, the discovery of neutrino oscillations (between flavor states $\nu_e, \nu_{\mu},$ and $\nu_{\tau}$)~\cite{Fukuda_1998, Ahmad_2002} necessitates that at least two neutrinos are massive particles. Each neutrino flavor thus consists of a superposition of three mass eigenstates, where the mixing components $U_{\alpha i}$ are elements of the 3 $\times$ 3 unitary Pontecorvo-Maki-Nakagawa-Sakata (PMNS) matrix~\cite{Pontecorvo:1957cp, Maki:1962mu}.\footnote{\color{black}For PMNS matrix element $U_{\alpha i}$, $\alpha \in \{e,\mu,\tau\} $ denoting flavor eigenstate, and $i \in \{1,2,3\}$ denoting mass eigenstate.} While the magnitudes of the squared differences between the three generations of neutrino mass eigenstates are known with high precision~\cite{PhysRevD.110.030001}, the absolute mass scale underpinning these values has yet to be measured.\footnote{\color{black}It additionally is unknown exactly what the ordering of these mass eigenstates (i.e., the neutrino mass hierarchy) is. The two possible cases, where the third mass eigenstate $m_3$ is the heaviest {\color{black}or} lightest of the three, are respectively referred to as Normal Ordering (NO) and Inverted Ordering (IO).} Neutrinos, in fact, are the only fundamental fermions whose masses remain unknown. Constraints nonetheless have been established on mass-relevant parameters from cosmological observations and various experiments. 

As the most abundant massive particle, neutrinos play a substantial role in the evolution of the universe and large-scale structure formation. Measurements of baryon acoustic oscillations from the Dark Energy Spectroscopic Instrument (DESI)~\cite{Levi2019Dark}, in aggregate with observational data from Planck~\cite{2020planck} and the Atacama Cosmology Telescope (ACT)~\cite{Madhavacheril_2024}, place the most stringent upper limit on the sum of neutrino mass eigenstates from cosmology: $\sum_{i=1}^3 m_i \leq$ {\color{black}72 meV} with 95\% confidence level~\cite{Adame_2025}. Inherent to this bound is the assumption of the standard $\Lambda$CDM cosmological framework~\cite{cosmol}, and the inclusion of model extensions or other nonstandard physics lessens the constraint (~\cite{Oldengott_2019, beyond}, for instance).

Additionally, neutrinos possess no electric or color charge, and have thus been posited to be their own antiparticle (a so-called ``Majorana particle'')~\cite{Majorana:1937vz}. Experiments have sought to test this {\color{black}hypothesis} by searching for neutrinoless double-$\beta$ decays~\cite{doublebeta} from various radioactive isotopes (such as $\ce{^{76}Ge}$~\cite{Gerda, Majoranadb}, $\ce{^{82}Se}$~\cite{CUPID}, $\ce{^{100}Mo}$~\cite{Amore, NEMO-3:2015jgm},  $\ce{^{130}Te}$~\cite{CUORE},  and $\ce{^{136}Xe}$~\cite{Kamland,EXO}, noninclusive). {\color{black}Outcomes of these searches set} an upper limit window on the effective Majorana mass $m_{\beta \beta}$.\footnote{ \color{black}$m_{\beta\beta}:=\sum_{i=1}^3 U_{ei}^2 m_i$.} The smallest of these windows was set by KamLAND-Zen~\cite{Kamland} at {\color{black}$m_{\beta \beta} < 36$ - $156 $ meV}, where the range is resultant from different nuclear calculations.

A more direct (and model-independent) {\color{black}method} to {\color{black}measure} the neutrino mass is to study the kinematics of electron capture~\cite{DERUJULA1982429} or $\beta$-decay as originally proposed by Enrico Fermi~\cite{Fermi}. The endpoint of the electron differential energy spectrum (i.e., maximal energy) in both cases is distorted directly according to the effective neutrino mass $m_\beta:=\sqrt{\sum_{i=1}^3 \left|U_{ei}\right|^2 m_i^2}$. While there are multiple ongoing efforts to study the electron capture kinematics of $\ce{^{163}Ho}$~\cite{Alpert_2015,Gastaldo_2014}, $\beta$-decay experiments have thus far proven {\color{black} to be the more prolific approach}.

{\color{black}O}ver the last 30 years, molecular tritium $\left(\ce{^{3}H_2}\right)$ $\beta$-decay experiments have been solely responsible for shrinking the world-leading upper limit on $m_\beta$~\cite{Formaggio_2021_trit}.} The current limit of $m_\beta$~<{\color{black}~450~meV with 90\% confidence level} was recently set by the Karlsruhe Tritium Neutrino (KATRIN) experiment~\cite{science}. {\color{black}The design of KATRIN, however, limits it to an idealized sensitivity of {\color{black}200~meV}~\cite{KATRINCollaborationKATRINCollaboration2005_270060419}, with background from surface radioimpurities~\cite{FRANKLE2022102686}, statistical uncertainty, and other systematic effects possibly resulting in a less-stringent final result~\cite{science}.  As a whole, molecular tritium experiments are inherently limited to a sensitivity of $m_\beta \approx$ {\color{black}100~meV} due to broadening caused by internal
molecular motion~\cite{PhysRevLett.84.242, PhysRevC.91.035505}.} Instead using {\it atomic} tritium $\left(\ce{^{3}H}\right)$, recently proposed projects such as PTOLEMY suggest that advances in our understanding of neutrino physics and techniques for high-sensitivity instrumentation could significantly improve sensitivity to measuring $m_\beta$~\cite{baracchini2018ptolemy,project8collaboration2022project}.

\subsection{The PTOLEMY Experiment}
\begin{figure}[!h]
\centering
    \includegraphics[width=0.9\linewidth]{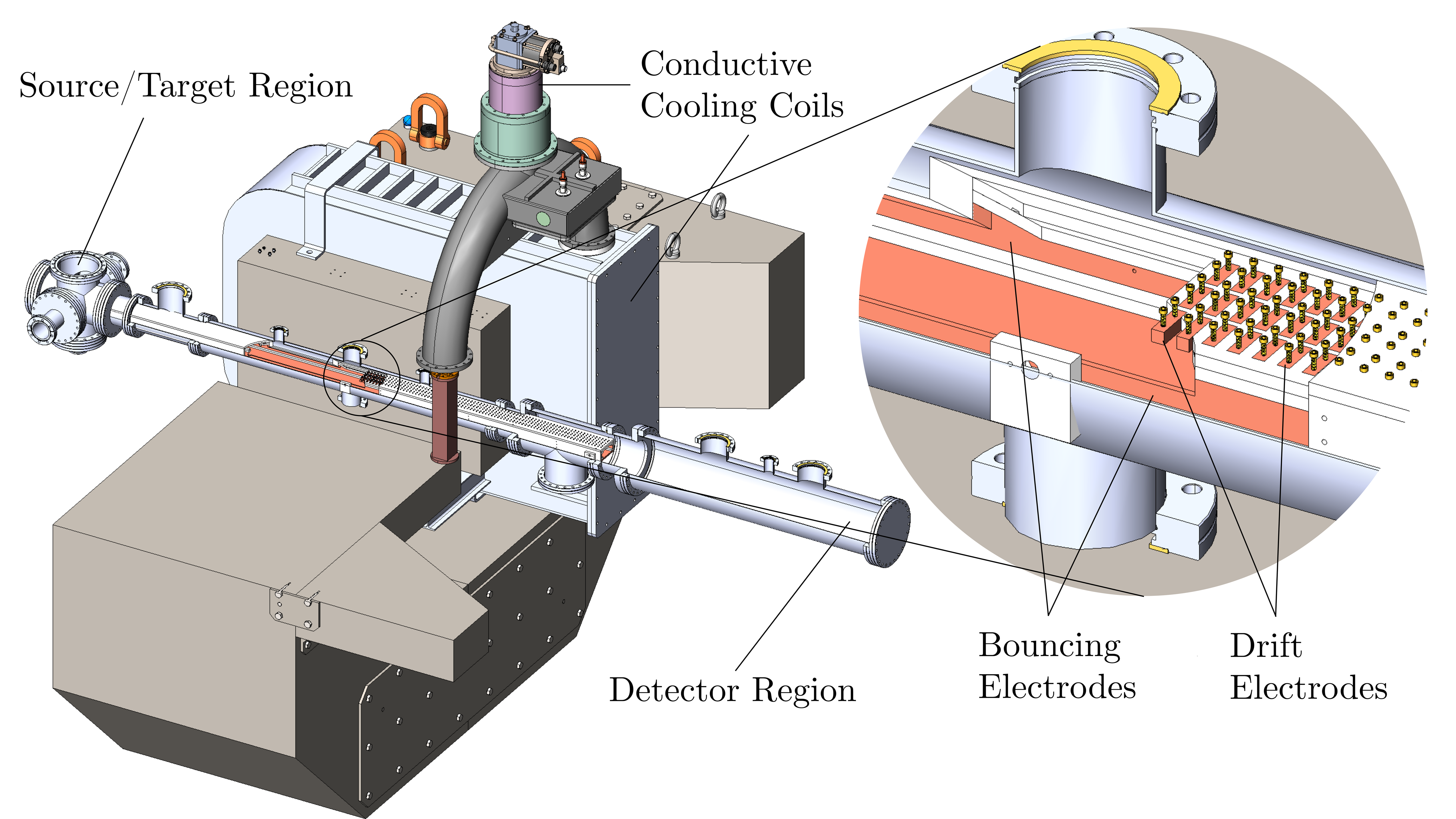}
    \caption{
    Schematic representation of the PTOLEMY prototype with conductive cooling MgB$_2$ coils. Electrons are emitted from the source/target region and enter the uniform ${\bf B}$-field region, where a transverse electric field induces an ${\bf E} \times {\bf B}$ drift. Within the highlighted section, RF tracking occurs in the uniform ${\bf B}$-field region between the pole faces, while transverse drift takes place in the decaying ${\bf B}$-field region, guided by the fringe field shaping extensions, gradually reducing the electrons’ energy. The bouncing electrodes provide additional control over electron trajectories in the longitudinal direction, ensuring efficient transport through the system. Finally, the electrons are directed toward the detector region, where their residual energy is precisely measured.
    }
    \label{fig:detect}
\end{figure}

PTOLEMY aims to create a scalable neutrino mass measurement experiment that performs precision spectroscopy of electrons produced near the endpoint energy from the $\beta$-decay of atomic tritium: {\color{black} $\ce{^{3}H} \to \ce{^{3}He} + e^{-} + \bar{\nu}_e $}. With strong background suppression and high energy resolution, distortions of the emission spectrum endpoint would enable a direct measurement of $m_\beta$. {\color{black} To achieve this, PTOLEMY proposes the following sequence: (1) guide tritium-emitted electrons into a uniform 1 T magnetic field region, (2) transmit them {\it slowly} through this uniform region via ${\bf E}$ $\times$ ${\bf B}$ drift (in the meanwhile using the emitted cyclotron radiation to estimate particle kinematics and select near-endpoint candidates), (3) transport selected candidates out of the uniform region and drastically reduce their kinetic energy by counterbalancing motion from ${\bf E}$ $\times$ ${\bf B}$ and ${\nabla {\rm B}}$ (where ${\bf E}$ is set according to the previously estimated kinematics), (4) and finally perform a highly precise measurement of the remaining kinetic energy}.\footnote{For a particle exposed to an electric field ${\bf E}$ while undergoing cyclotron motion due to a magnetic field ${\bf B}$ (satisfying $E^2 < c^2 B^2$), the \emph{drift velocity} of its guiding center (the center point of circular cyclotron motion) is given as ${\bf v}_{\rm E} = \frac{{\bf E \;} \times {\;\bf B}}{\rm B^2}$ and commonly referred to as ${\bf E}$ $\times$ ${\bf B}$ drift. {\color{black}For a non-uniform magnetic field, the total net force is given as ${\bf f} = -\mu \nabla {\rm B}$, and a similar drift term arises: ${\bf v}_{\nabla {\rm B}} = \frac{ T_\perp}{q {\rm B}}\frac{{\bf B \;} \times \; \nabla{\rm B }}{{\rm B}^2}$, where $\mu$ is the orbital magnetic moment, $q$ is charge, and $ T_\perp$ is the kinetic energy component of the particle transverse to the magnetic field direction.}} A schematic of the PTOLEMY prototype is shown in Figure~\ref{fig:detect}.

Tritium was selected for multiple reasons, including {\color{black} suitable half-life ($\approx 12 .3$ years)} and low $Q$-value ($\approx 18.6$ keV)~\cite{Cocco_2008}. The proposed target substrate is graphene~\cite{Betti_2019}, which can efficiently store atomic tritium by binding it to carbon atoms. Extensive work is currently being undertaken by the collaboration and community at large to study the solid state effects on the $\beta$ emission spectra~\cite{PhysRevD.104.116004, PhysRevD.105.043502,10.21468/SciPostPhys.17.1.022, PhysRevD.106.053002, casale2025betadecayspectrumtritiatedgraphene}.

 To attain high measurement resolution, PTOLEMY utilizes transition-edge sensor (TES) microcalorimeters, which precisely measure energy via the strongly temperature-dependent resistance of a cryogenic superconducting material's phase transition. In order for tritium endpoint electrons to fall within the dynamical range of the TES, static electromagnetic fields are used to filter their kinetic energy from 18.6 keV to roughly 10 eV. This is performed by PTOLEMY's novel \emph{transverse drift filter}~\cite{2Betti_2019}. The filter uniquely orients both the magnetic field gradient and electric field orthogonal to the magnetic field, resulting in components of the electron's motion driven by $\nabla {\rm B}$ and ${\bf E}$ $\times$ ${\bf B}$. {\color{black} The force produced by $\nabla {\rm B}$, when accompanied by {\bf E} $\times$ {\bf B} drift,  can do work on an electron and reduce its kinetic energy transverse to the magnetic field ($T_\perp$). Namely, $\frac{dT_\perp}{dt} = \frac{\mu}{\rm B^2} {\bf E} \cdot (\nabla {\rm B} \times {\bf B})$, indicating a change in voltage potential corresponds to a change in kinetic energy. Hence, a particle steadily loses kinetic energy as it climbs a higher potential barrier through the filter and travels towards the TES. Applying an additional electric field parallel to the magnetic field facilitates a reduction in parallel kinetic energy ($T_\parallel$) and thus total kinetic energy ($T = T_\parallel + T_\perp$).} The design of the transverse filter allows for a compact, scalable detector, with each dimension on the order of 1 meter~\cite{Apponi_2022}. In total, this technique could achieve an energy resolution as small as {\color{black}50 meV}~\cite{Rajteri:2019eua}, opening the possibility to a detection of the Cosmic Neutrino Background~\cite{PhysRev.128.1457,Cocco_2008} in addition to an effective neutrino mass measurement.

{\color{black} The drawback, however, is that a particle's emission angle must be accurately known to properly tailor an orthogonal electric field in the filter that depletes $T_\perp$ (i.e., while near-endpoint electrons have similar energies, their momenta split with respect to the magnetic field can vary drastically). Further, this field necessarily must be set {\it prior} to the particle entering the filter to maintain electrostatic conditions. To accurately infer such emission spectra, }PTOLEMY {\color{black}utilizes} \emph{cyclotron radiation emission spectroscopy} (CRES), which {\color{black}is a technique} pioneered by the Project 8 collaboration~\cite{PhysRevD.80.051301,PhysRevLett.114.162501}.  Using radio-frequency (RF) antennae, an electron's {\color{black}total and perpendicular kinetic energy are estimated by the relativistic frequency shift of the emitted cyclotron radiation {\color{black}as} it transits through a uniform 1 T magnetic field ``RF region'' (preceding the transverse drift filter) via ${\bf E}$ $\times$ ${\bf B}$ drift. The RF region consists of electrostatic mirrors parallel to the magnet pole faces (to trap the particle) and an orthogonal electric field set according to voltage potentials. Particularly, ${\bf E}$ $\times$ ${\bf B}$ drift is precisely slowed to trap electrons in the region long enough to obtain an adequate integrated radiated power signal for the CRES measurement. In the instance of an inferred near-endpoint electron, the $T_\perp$ estimate is used to adjust voltage potentials, producing a suitable electric field in the drift filter.} Provided enough time to obtain a large cyclotron radiation signal, CRES has been shown to stringently limit backgrounds near the tritium endpoint~\cite{PhysRevLett.131.102502}, 
enabling the possibility of direct differential energy measurement in time coincidence following a non-destructive RF tag.  With an energy resolution of $\mathcal{O}(1 \:{\rm eV})$, Project 8 has utilized CRES to establish a neutrino mass sensitivity of $m_\beta \lesssim$ {\color{black}155 eV}~\cite{PhysRevLett.131.102502}. Addressing plans and challenges to scale up the experiment and implement an atomic tritium source, they propose to achieve a sensitivity of $m_\beta~\lesssim$~{\color{black}40~meV}~\cite{Project8:2017nal}. Instead, PTOLEMY uses CRES solely to obtain an estimate of the particle's kinematics with substantially coarser resolution than the final TES measurement.

The collaboration has performed CRES measurements of conversion electrons emitted from $\ce{^{83m}Kr}$ gas contained in an electromagnetic trap, successfully identifying signal tracks akin to observations by Project 8~\cite{Pesce}; a publication of these results is expected soon.
In addition, detailed simulation work has been undertaken to understand RF emission and antenna response in a system similar to the future RF region, including estimates of required electron time of flight and the development of a fitting algorithm to extract kinematic parameters~\cite{iwasaki2024cresbasednondestructiveelectronmomentum}. However, electron ${\bf E}$~$\times$~${\bf B}$ transport as outlined for the RF region has yet to be shown in a physical setup. This is a crucial milestone in actualizing the detector, {\color{black}enabling} future CRES measurements, filter implementation, and source injection.

\subsection{\boldmath ${\bf E}$ $\times$ ${\bf B}$ Transport}
${\bf E}$ $\times$ ${\bf B}$ drift has been extensively researched in the contexts of plasma dynamics, propulsion, and astrophysical phenomena (for instance:~\cite{Osorio_Quiroga_2023},~\cite{doi:10.2514/6.2005-4405}, and~\cite{10.1093/mnras/stab2739}), though is seldom employed deliberately for charged particle transport. {\color{black}Notwithstanding, it is relatively straightforward to produce and implement, especially in setups already involving magnetic fields. The addition of electrodes in a Penning ion source test stand has been presented as a means to trap secondary electrons by {\bf E}~$\times$~{\bf B} drift~\cite{Savard_2022}, and KATRIN has shown the use of non-parallel electric and magnetic fields at the emission point of a photoelectron gun yields an apt angular-selective calibration source for their experiment~\cite{KValerius_2011}. Further yet, {\bf E} $\times$ {\bf B} drift, in conjunction with an electrostatic barrier (i.e., the concepts utilized for transport in PTOLEMY's RF region), has been implemented for positron injection into a magnetic dipole trap~\cite{10.1063/5.0050881}. These prior studies, however, have no consideration of particle time of flight with respect to {\bf E} $\times$ {\bf B} drift, nor do they manipulate electric fields to slow transit speed.\footnote{\color{black}By time of flight, we are referring to the amount of time a particle takes to travel through a defined distance or volume.} In this regard,} PTOLEMY's concept of \emph{slow} ${\bf E}$ $\times$ ${\bf B}$ drift for low-energy electron transport is entirely novel. Tethered to the strength of ${\bf E}$, an electron's drift velocity can easily be changed by adjusting voltage potentials throughout the system. This feature is crucial to precisely reduce drift velocity near the RF antennas such that there is enough time (on the order of 100 $\mu$s~\cite{iwasaki2024cresbasednondestructiveelectronmomentum}) to obtain a large enough signal to accurately tag electrons.

\section{${\bf E} \times {\bf B}$ Demonstrator Design}

{\color{black}
\subsection{Aims and Scope} \label {sec:aims}
This paper presents a working physical setup that transports $\beta$-decay electrons as described for the PTOLEMY RF region. Namely, an emitted electron is maintained within a defined volume via electrostatic mirrors and transits by {\bf E} $\times$ {\bf B} drift. In the central portion of this volume, ${\bf E}$~$\times$~${\bf B}$ drift is reduced to mimic an injected particle entering the RF region, slowly passing through, then subsequently regaining drift speed as it approaches the filter. CRES measurements are not performed here; rather, transported electrons are determined according to a detector placed on the other side of the slow drift region. The recorded information is used to tabulate transmission rates.

Though we are primarily interested in maximally slowing the electron in the physical setup, a more detailed characterization is necessary to undergird a correspondence to simulation, thereby validating its function. With a programmable high voltage power supply, {\color{black}we} record event rates under varied voltage potential configurations. We thus develop frameworks to systematically alter voltages and create coherent event rate datasets, which we call `Modes of Operation' and describe further in Section~\ref{modesss}. A correspondence between these datasets and their simulation counterparts confirms system function and allows us to extract time of flight information.

Importantly, this work intends to show the {\it feasibility} of actualizing slowed {\bf E} $\times$ {\bf B} drift, including practices to facilitate the RF tagging capability. It {\color{black}does not serve} as a high-precision instrument or is reflective of the future capabilities of the PTOLEMY RF region, {\color{black}but rather exhibits a novel technique in a simple setup.} We conclude this section by discussing some perceived limitations.
\begin{enumerate}
    \item {\bf Attainable Time of Flight}: To emulate the PTOLEMY RF region, we wish to observe slowed ${\bf E}$ $\times$ ${\bf B}$ drift in a {\it uniform magnetic field}. Unfortunately, the uniform field volume of the electromagnet available for this study is fairly limited $\left(\approx 190 {\: \rm mm} \times 150 {\: \rm mm} \times 90{\: \rm mm}\right)$. This directly restricts the dimensions of the setup and, hence, electron time of flight. The {\color{black}situation} is exacerbated {\color{black}due to a} persistent residual electric field in the slow drift region (produced from other voltage potentials), which results in a {\color{black}sizable} finite minimum field. We are additionally constrained (for reasons discussed in Section~\ref{layoutandmagnet}) to operate at a magnetic field of only 160 mT. In conjunction, this places a strict lower bound on achievable drift speed.
    
    Though we show an $\mathcal{O}(1)$ manipulation of drift speed, the slowest attainable time of flight remains orders of magnitude smaller than the future needs of the detector ($\approx 60$ ns, {\color{black}as compared to the} previously quoted 100 $\mu$s). We are confident, however, that significantly slower drift can be achieved with a larger setup and stronger magnetic field, enabling an electron containment of 100 $\mu$s if not longer.
    \item {\bf\boldmath${\beta}$ Electron Source}: For evident reasons of cost and handling, we choose to use a different radioactive isotope as a $\beta$ electron source in place of $\ce{^{3}H}$. We selected a $\ce{^{14}C}$ source for its comparable $Q$-value, vacuum hardiness, and availability at our institution.
    \item {\bf Electron Detection}: The primary function of a detector for this setup is to properly detect the transmitted $\beta$ spectrum above a minimum-energy threshold. A precise, high-resolution instrument tailored for low-energy electron detection would be superfluous for our needs. Under additional cost, size, and field resilience considerations, we opt for a commercial photodetector. With slight alterations, it is sensitive to a significant portion of the $\ce{^{14}C}$ $\beta$ spectrum. Details are provided in Section~\ref{PINDIODE}.

     The detector is intended to serve as a broad tag for electrons, in which case details of resolution and systematics are less important for the scope of this study. A {\color{black}comprehensive} energy measurement would require full calibration, establishing an electron energy resolution (preferably with multiple mono-energetic sources), an evaluation of dead layer thickness (not provided by the manufacturer) and resultant average energy loss based on particle impingement angle, a meticulous assessment of the minor output changes under varied electric and magnetic fields, an understanding of changing output response due to self-heating, an estimation of electron energy loss due to cyclotron radiation emission and imperfect vacuum, and other {\color{black}possible} systematic effects (such as noise contributions from other devices). While KATRIN has successfully characterized and employed a very similar detector for high-resolution measurements~\cite{WUSTLING2006382}, the sole requirement for our system is to determine a minimum energy detector sensitivity (to properly compare with simulation). We resolve this {\color{black}by} using a beta spectrometer, as discussed in the Supplemtary Material. With no available information regarding energy resolution, we approximate it with the manufacturer-provided photocurrent uncertainty (11\%).

\item {\bf Systematic Uncertainties}: We are confident in system operation by identifying corresponding, broad trends between recorded data and simulation (manifesting as a least-squares regression fit). While important, a rigorous examination of systematic uncertainties is not necessary to achieve the outlined goal of this {\color{black}demonstration}: feasibility, {\it not} precision. For consideration of future implementation, we discuss various contributions to uncertainty. However, we estimate uncertainty by assessing the fluctuation of repeated measurements, which we consider suitable for our purposes.
\end{enumerate}

}
\subsection{Concept} \label{sec:concept}

\begin{figure}[!h]
\centering
    \includegraphics[width=0.7\linewidth]{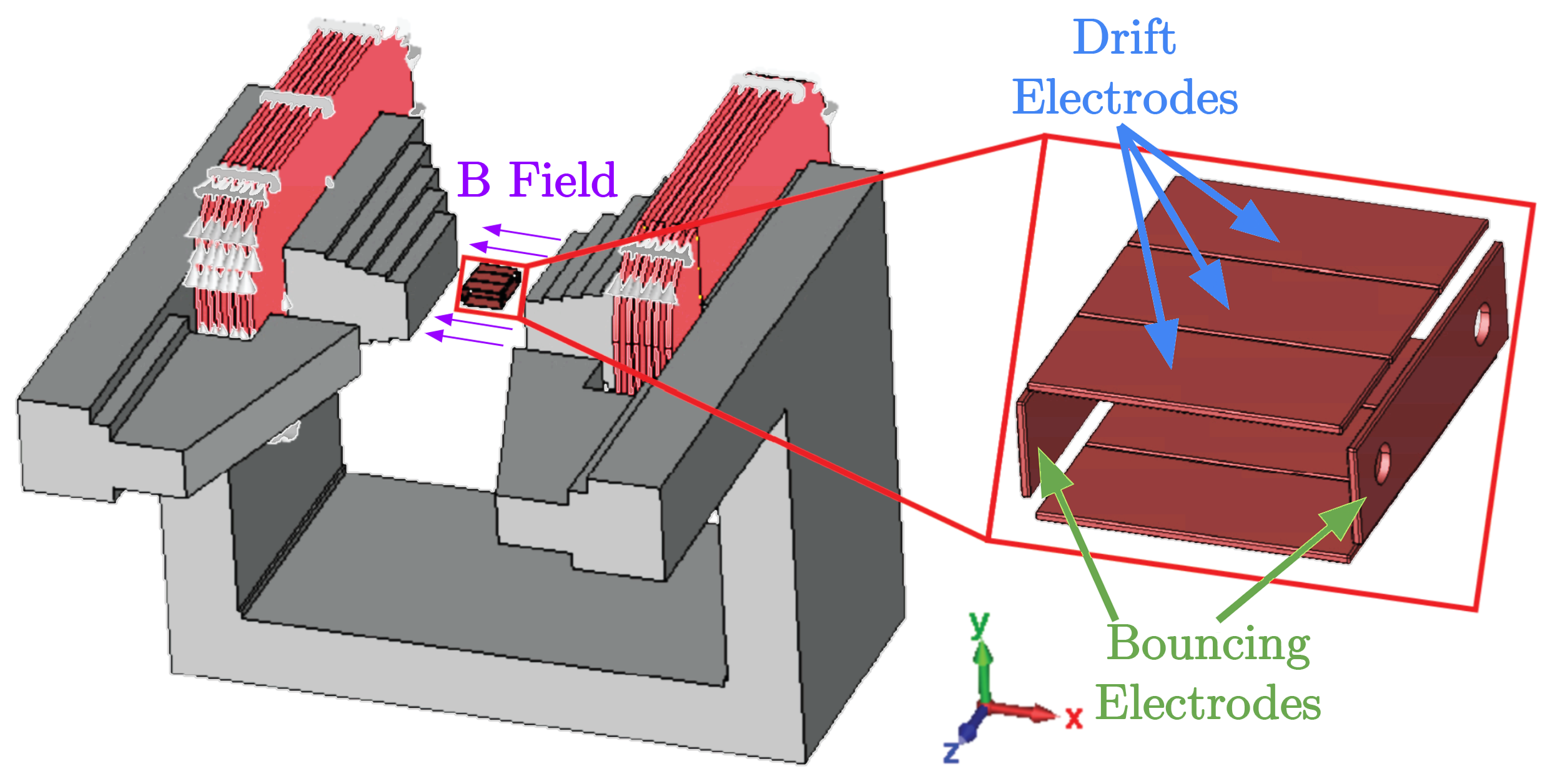}
    \caption{Conceptual demonstrator setup. Electrode plates are positioned between a magnet's pole faces, where a uniform magnetic field is generated (shown in purple). The plates (enlarged on the right) consist of two bouncing electrodes (green) and six drift electrodes (blue). The bouncing electrodes are oriented parallel to the magnet pole faces and supplied with a negative potential to contain electrons within the system. The six drift electrodes are organized into three vertically offset pairs that produce an electric field orthogonal to the magnetic field. The potential difference in the center drift plates is set to be smaller than the outer plates, facilitating slower ${\bf E}$ $\times$ ${\bf B}$ drift. Following the included axes, the magnetic field is negatively oriented along the x-direction and the electric field (produced by the drift electrodes) is oriented along the y-direction, resulting in ${\bf E}$ $\times$ ${\bf B}$ drift in the z-direction.}
    \label{fig:setup}
\end{figure}

\begin{figure}[!h]
\centering
    \includegraphics[width=0.6\linewidth]{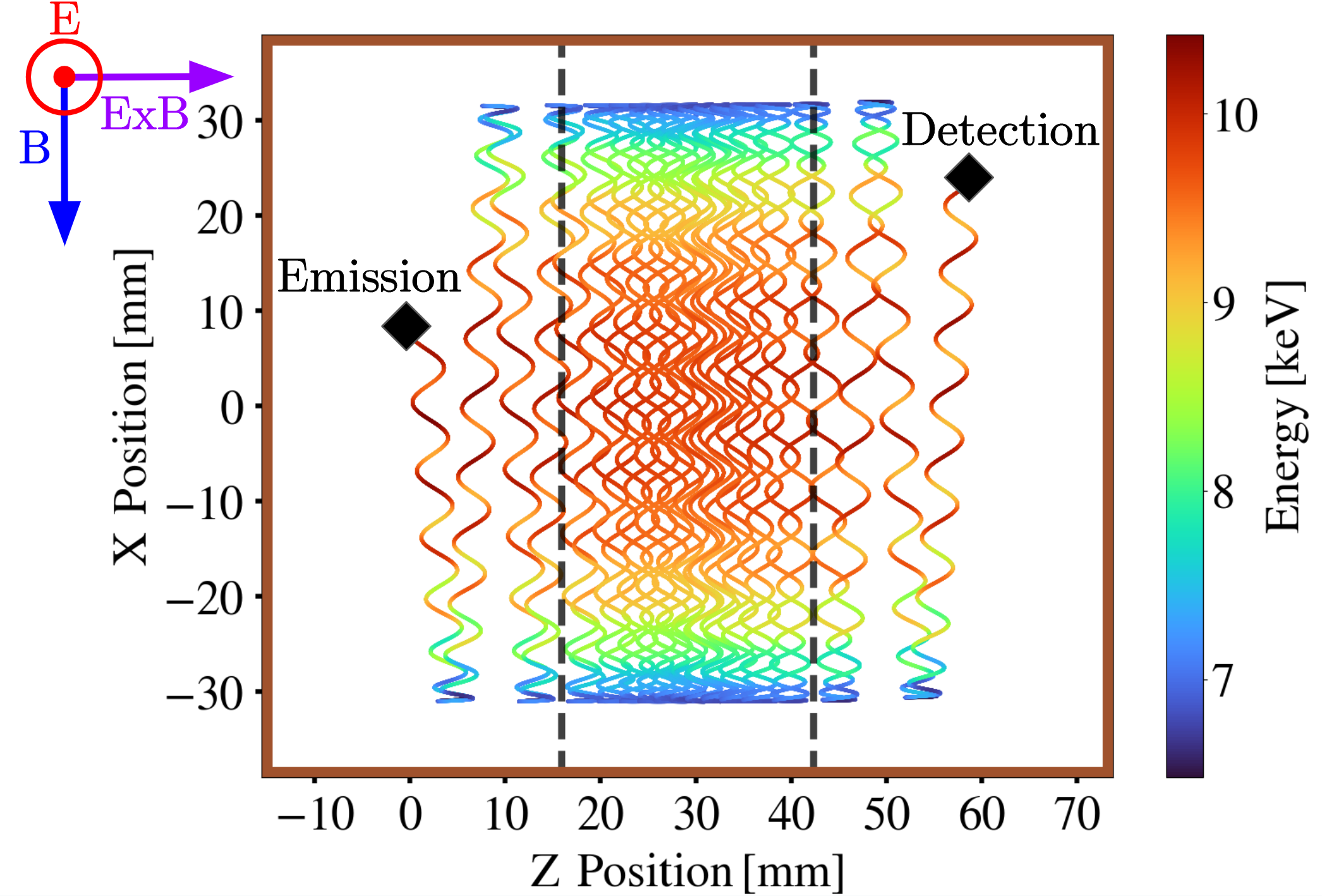}
    \caption{Top-down trajectory of a 10 keV electron within the demonstrator. The electron exhibits cyclotron motion while bouncing between two electrodes and drifting rightward due to potential differences in the other six. A smaller difference on the center pair of drift electrodes (marked by dashed vertical lines) results in slowed drift, as evidenced by an increased density of trajectory lines. Emission and detection points of the electron are denoted with black diamonds. The direction of the electric field (red), magnetic field (blue), and ${\bf E}$ $\times$ ${\bf B}$ drift (purple) are included in the upper left corner. The brown outline represents the outer perimeter of the electrode plates. In total, emitted electrons drift approximately 6 cm.}
    \label{fig:traject1}
\end{figure}

The conceptual {\color{black}design for the ${\bf E}$ $\times$ ${\bf B}$} demonstrator is shown in Figure~\ref{fig:setup}. Eight electrode plates are situated between the pole faces of a magnet. Two of these electrodes, the \emph{bouncing electrodes}, are oriented parallel to the magnet pole faces and supplied with a negative voltage to create a potential barrier to contain electrons within the apparatus (such that the electron motion parallel to the magnetic field exhibits a \emph{bouncing} behavior), similar to the electrostatic mirror described in~\cite{10.1063/1.5097388}. The remaining six are organized into three pairs of vertically offset \emph{drift electrodes}, with voltages applied to generate an electric field orthogonal to the magnetic field. The potential difference between the center pair of plates is set to be smaller than that of the outer pairs, thereby creating a region of slower drift. In total, an electron injected through one end of the demonstrator undergoes cyclotron motion, bounces between bouncing electrodes, and drifts toward the other end. An example of such a trajectory is shown from a top-down orientation in Figure~\ref{fig:traject1}. The increased density of trajectory lines in the center region indicates slower drift. {\color{black}No work is performed by ${\bf E}$ $\times$ ${\bf B}$ drift, hence electron kinetic energy only deviates in accordance with the potential experienced from the bouncing electrodes throughout the trajectory.} {\color{black} As with the physical setup, the detection location in this simulation is maintained at a voltage potential comparable to the emission point ($\approx 0$ V), ensuring that the final electron kinetic energy roughly coincides with that originally emitted.} This Figure and all subsequent simulation estimates are generated using CST Studio Suite by Dassault Syst\`emes~\cite{CSTStudioSuite}.

\subsection{Field Cage}
\begin{figure}[!h]
\centering
    \includegraphics[width=.85\linewidth]{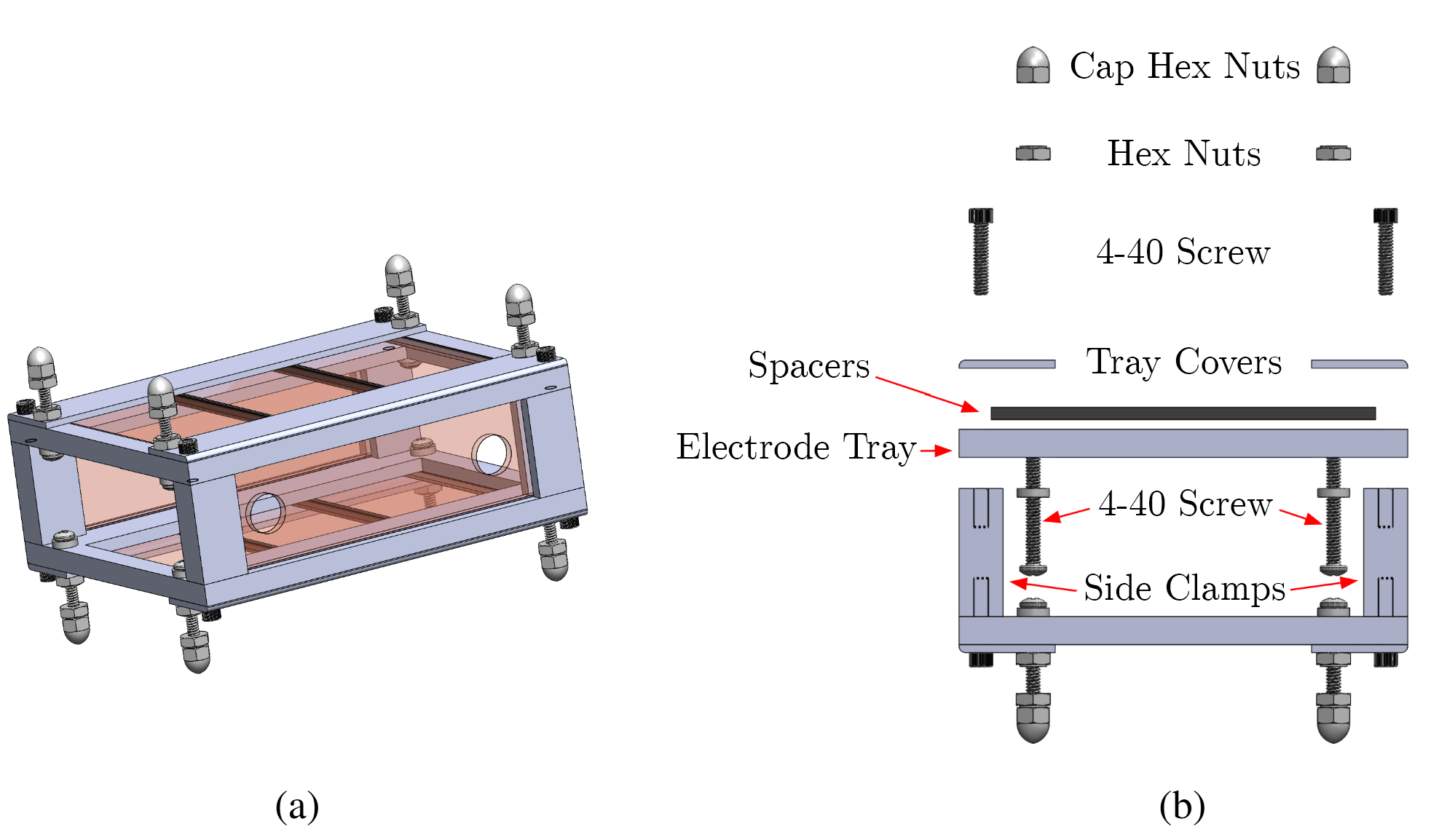}
    \caption{(a): Field cage design to fix the relative position of electrode plates. Components consist of machined Teflon pieces and commercially available plastic screws, washers, and hex nuts. (b): Exploded view of field cage with labeled parts. Bouncing electrodes are held via clamps, and the trios of drift electrodes lay on a tray fixed by a screwed-in cover. Drift electrodes are separated by 4 mm spacers to prevent electrical shorting, and the outward-facing screws and cap hex nuts are used to fix the field cage in a vacuum chamber.}
    \label{fig:thecage1}
\end{figure}

\begin{figure}[!h]
\centering
    \includegraphics[width=0.5\linewidth]{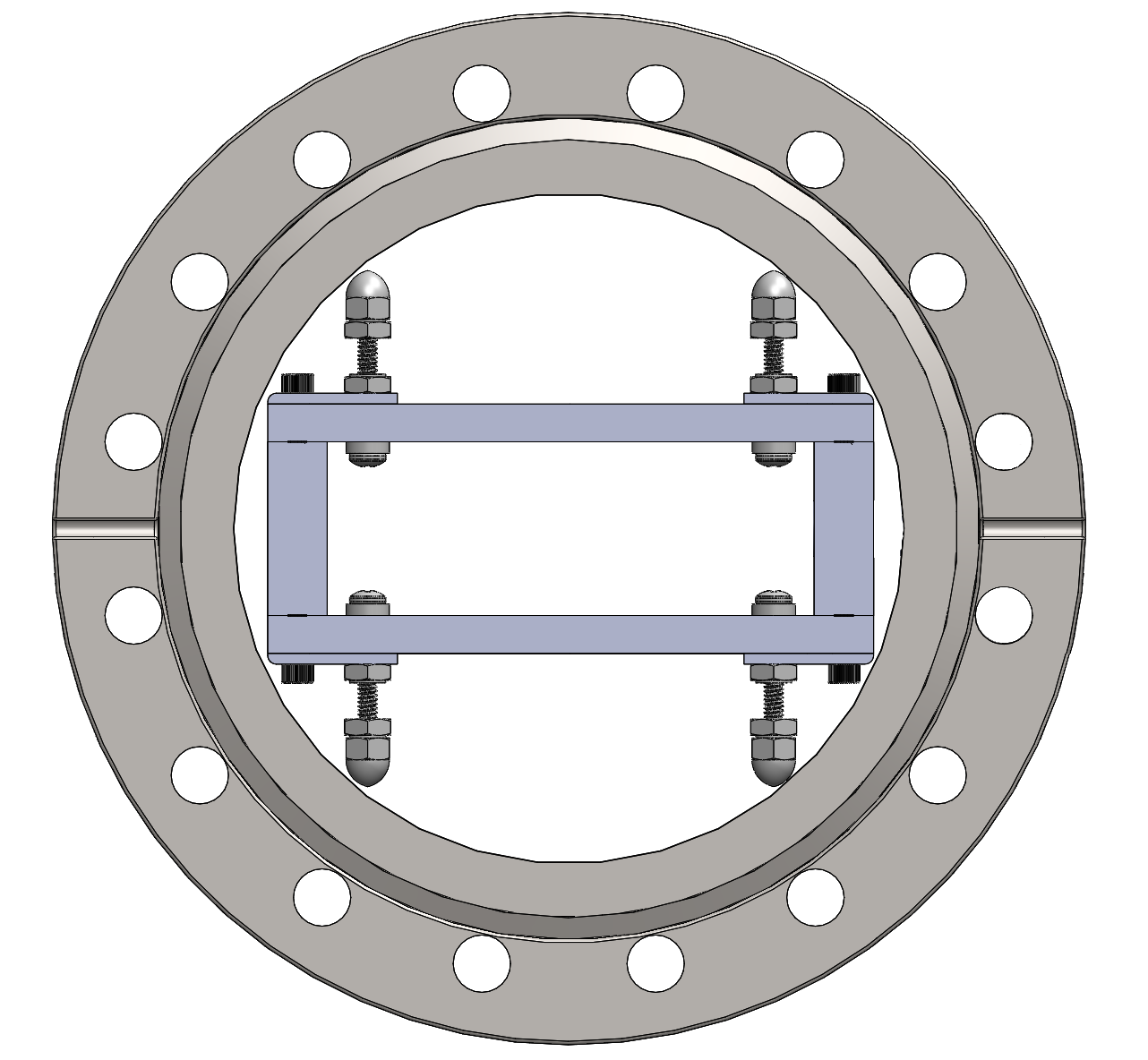}
    \caption{Side profile of the field cage situated in a $10.2{\: \rm cm}$ OD vacuum chamber. The cap hex nuts are adjusted to prevent movement of the field cage while in operation.}
    \label{fig:sidenip}
\end{figure}

At the forefront of actualizing the design shown in Section~\ref{sec:concept} is a \emph{field cage} to hold the electrodes. Assembled with machined Teflon parts and plastic screws, the custom-built cage fixes the relative positions of all eight copper electrode plates. Teflon was chosen for its excellent insulation properties and quick outgassing in vacuum. Two bouncing electrodes $\left(96.8 {\: \rm mm} \times 25.4 {\: \rm mm} \times 2.4{\: \rm mm}\right)$ span the length of the cage on both sides. Two holes $\left(10.2 {\: \rm mm} \: {\rm diameter} \right)$ were bored into one of these electrodes originally for electron injection/detection, though are utilized to spatially fix additional components in the final setup. The outer $\left(76.2{\: \rm mm} \times 31.8{\: \rm mm} \times 2.4{\: \rm mm}\right)$ and slightly thinner central $\left(76.2{\: \rm mm} \times 25.4{\: \rm mm} \times 2.4{\: \rm mm}\right)$ drift plates are separated by plastic spacers of thickness $4{\: \rm mm}$ to prevent electrical sparking~\cite{zap}. These plates lay flat on a tray and are fixed by a cover that screws into the bouncing electrode clamps. The upper and lower trio of plates are separated vertically by $31.8{\: \rm mm}$. The design and an `exploded' view of the casing including labeled parts are shown in Figure~\ref{fig:thecage1}. The outward-facing Polyetheretherketone (PEEK) screws and cap hex nuts ensure the field cage remains fixed when situated inside a $10.2{\: \rm cm}$ outer diameter (OD) vacuum chamber as shown in Figure~\ref{fig:sidenip}.

\subsection{\boldmath${\beta}$ Emission Source}

\begin{figure}[!h]
\centering
    \includegraphics[width=0.55\linewidth]{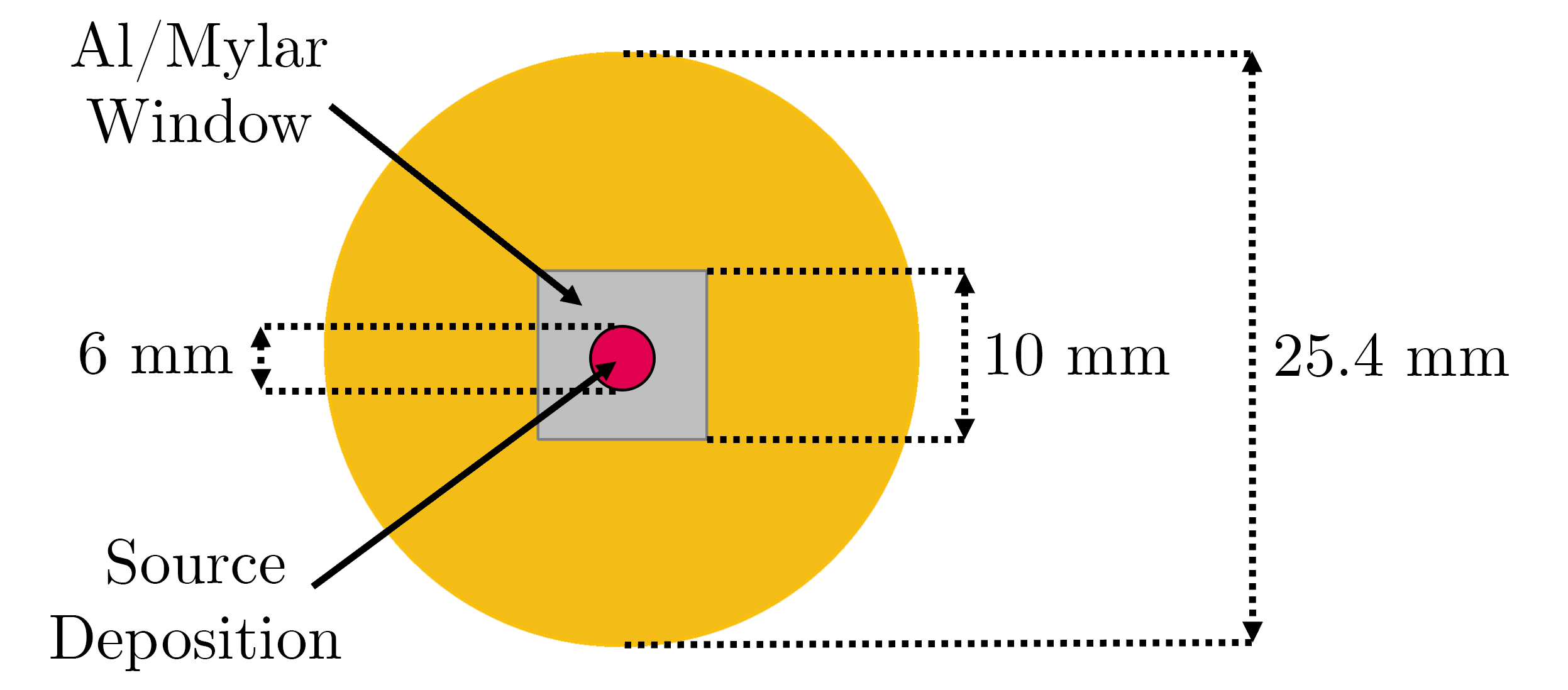}
    \caption{Schematic of $\ce{^{14}C}$ disk used as $\beta$ source in this study. The red dot indicates the 10 $\mu$Ci deposition of the radioactive source, and the grey square represents the aluminized Mylar window. $\ce{^{14}C}$ was selected for its relatively similar Q-value to tritium. The design of the source disk allows for it to safely withstand vacuum-level pressures.}
    \label{fig:1a}
\end{figure}

$\ce{^{14}C}$ was selected as a suitable beta emitter for its relatively similar $Q$-value (155 keV) to tritium. The specific source used is a 10 $\mu$Ci $\ce{^{14}C}$ deposition with an aluminized Mylar window on a $25.4{\: \rm mm}$ diameter laminate disk manufactured by Spectrum Techniques. As $\ce{^{14}C}$ is a sole beta emitter, an activity of 10 $ \mu$Ci corresponds to an emission rate of 370,000 electrons per second. The source disk ($2\pi$ solid angle) emits half of that (i.e., 185,000 electrons per second), which was verified with a Geiger counter. The aluminized Mylar window allows the source to safely withstand vacuum-level pressures. A schematic is shown in Figure~\ref{fig:1a}.

\subsection{PIN Diode Detector}\label{PINDIODE}
\begin{figure}[!h]
\centering
    \includegraphics[width=0.85\linewidth]{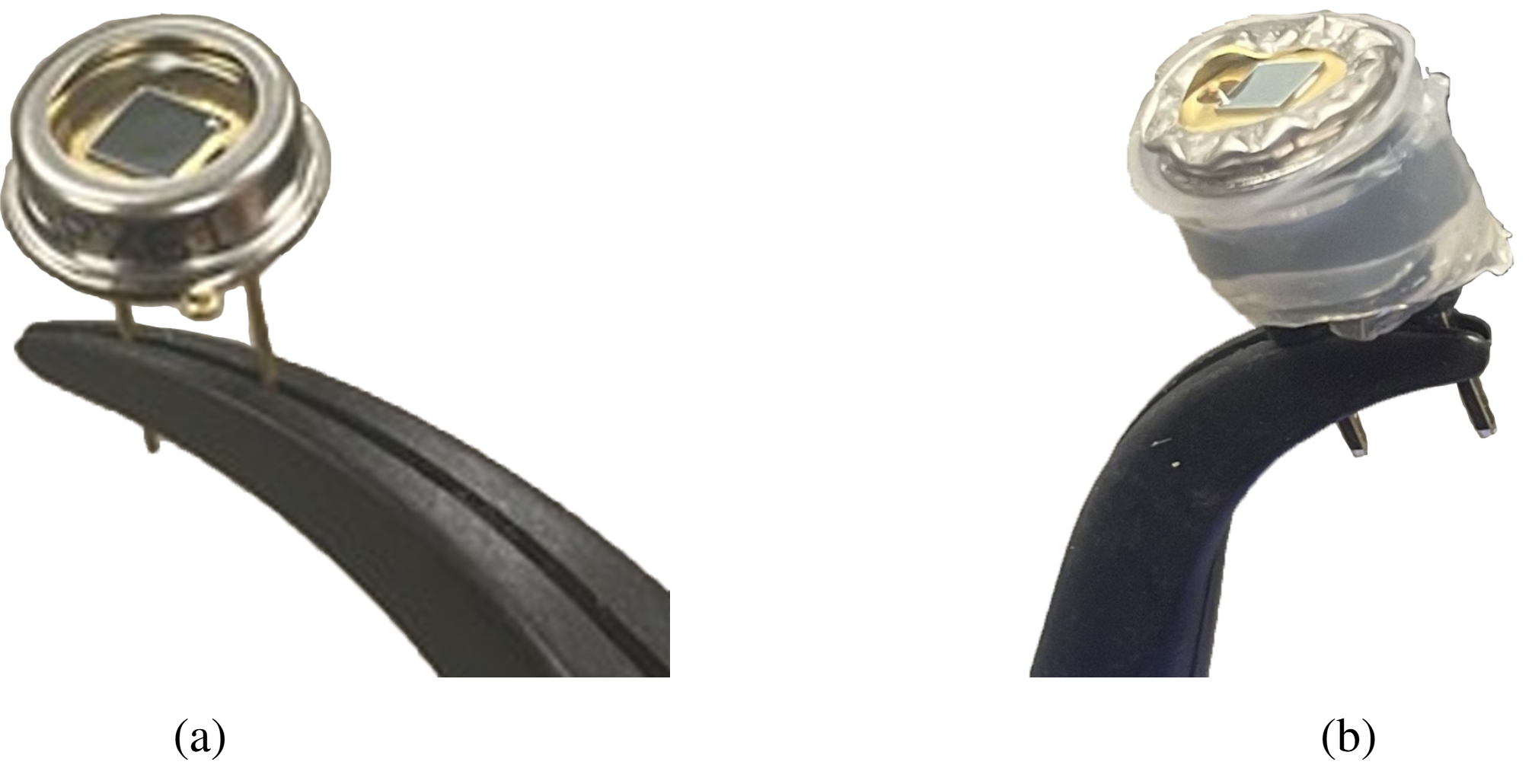}
    \caption{BPX 61 diode (a) unaltered and (b) altered with removed glass cover, flattened casing, and heat shrink tube wrapping. This diode was selected as our detector for its commercial availability and low cost. Alterations to the device were performed to enable low-energy electron detection in a high magnetic field.}
    \label{fig:diode_both}
\end{figure}

For electron identification, the system size and operating environment necessitate the use of a detector that is both compact and resilient to high magnetic fields. Inspired by the utilization of PIN diodes for low-energy electron detection by KATRIN~\cite{WUSTLING2006382}, the OSRAM BPX 61 Silicon PIN diode was selected as a detector for this study. The specific model was chosen for its commercial availability and comparatively low cost. PIN diodes are electrical components containing an undoped intrinsic semiconductor region between a p-type and n-type semiconductor. When operated in reverse bias, they are sensitive to energy depositions of charged particles and light. The BPX 61 additionally features a low dark current of 2 nA and a relatively large active detection area ($7.02 \:{\rm mm}^2$), limiting electrical noise while increasing possible event count.
However, commercial diodes --- including the BPX 61 --- are typically hermetically sealed by plastic or glass to prevent damage to the semiconductor region. Such covers, though providing little impedance to light, have a great enough stopping power to entirely prevent low-energy electron detection. For our study, the cover was removed and the exterior casing was flattened to the base of the diode, improving sensitivity to higher-energy electrons with larger cyclotron radii.  A heat shrink tube was attached around the neck of the diode to prevent torsion/bending from magnetization. An image of the BPX 61 diode and an example of the performed alterations are shown in Figures~\ref{fig:diode_both}.

\begin{figure}[!h]
\centering
    \includegraphics[width=0.5\linewidth]{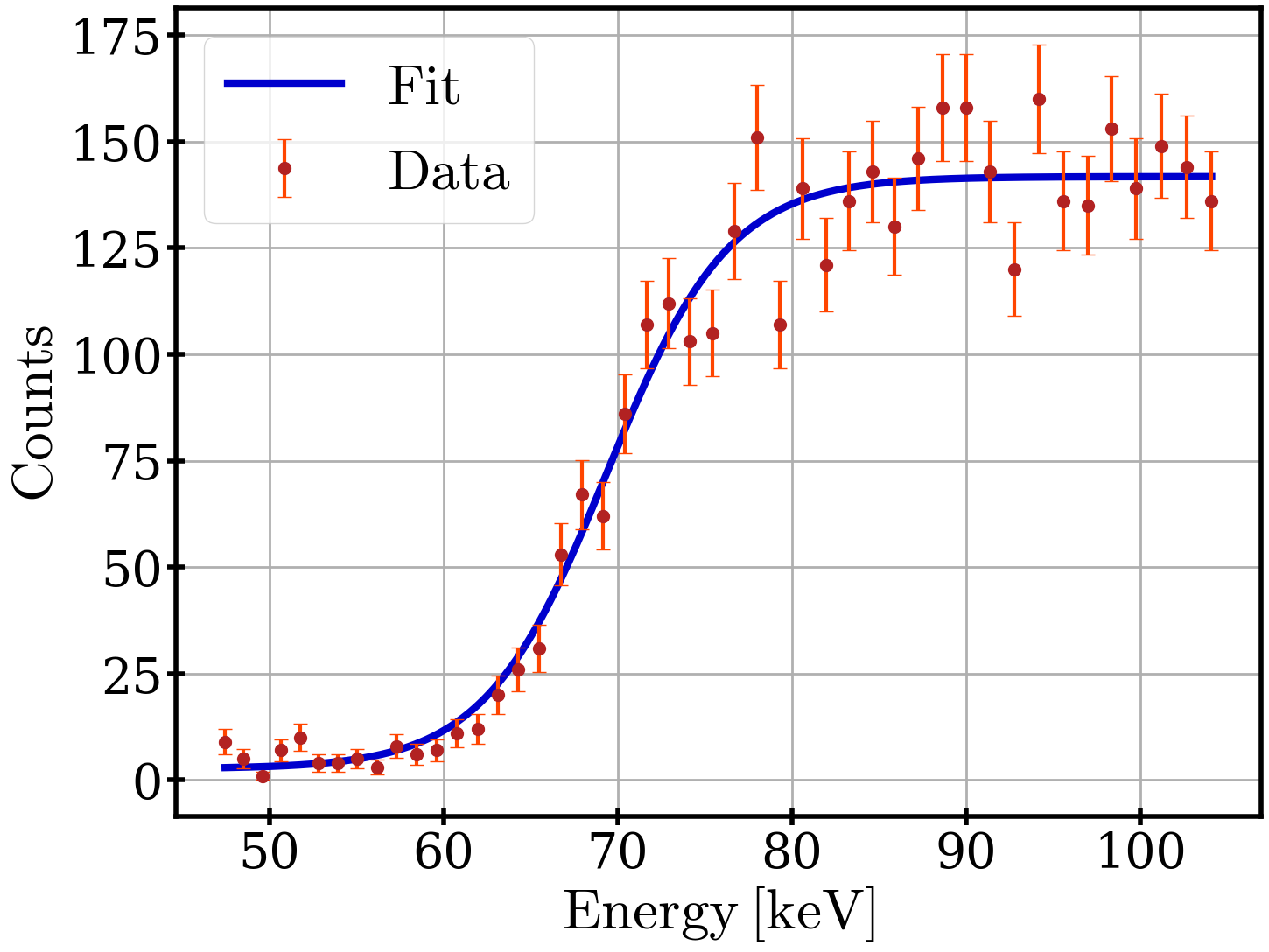}
    \caption{Diode energy threshold characterization using a beta spectrometer. Shown are counts of recorded events (with Poisson statistic error bars) as a function of emitted particle energy, as well {\color{black}as a fit to the data to obtain an overall trigger count curve.} A half-maximal trigger count is seen to occur at roughly 70 keV.}
    \label{fig:diode_thresh}
\end{figure}

Energy threshold characterization of the device was performed using a beta spectrometer (details provided in Supplementary Material). {\color{black}In summary, the spectrometer uses a solenoidal magnetic lensing system to generate a cylindrically-symmetric magnetic field. Emitted electrons undergoing a helical trajectory require a certain momentum (and hence energy) to pass circular barriers inside the device before reaching the detector. The spectrometer sequentially selects ranges of allowable momenta with a constant resolution of 2.45\% ($\sigma/\mu$) by incrementally increasing the strength of the solenoidal magnetic field. Converting momentum to energy, each increment corresponds to a specific source energy range where events are recorded by the PIN diode. Each setting lasts roughly 1 minute, and the full range of settings is cycled through on the order of 50 times (to account for time-dependent noise).} Events are considered to be waveforms whose peak height exceeds a value of 45 mV above baseline after amplification (the electronic and amplification chain are identical to those discussed in Section~\ref{electronics}). Based on the phase space of allowed particle trajectories in the system (discussed in Section~\ref{simulations}), we situate the diode at a shallow angle ($\approx 20\degree$) with respect to the instrument aperture. In Figure~\ref{fig:diode_thresh}, we present a trigger efficiency curve for the device obtained from beta spectrometer counts, observing a half-maximal trigger count occurring at $\approx 70$ keV. {\color{black}This curve is then scaled to obtain a relative trigger efficiency.} To clarify, this result does not reflect the true sensitivity of the device, but rather the trigger efficiency at a threshold safely above the noise floor of our subsequent data-taking conditions. Following manufacturer specifications {\color{black}on photocurrent uncertainty}, we assign an energy resolution of 11\%.   

\subsection{Electronics and Readout Chain}\label{electronics}
\begin{figure}[!h]
\centering
    \includegraphics[width=0.45\linewidth]{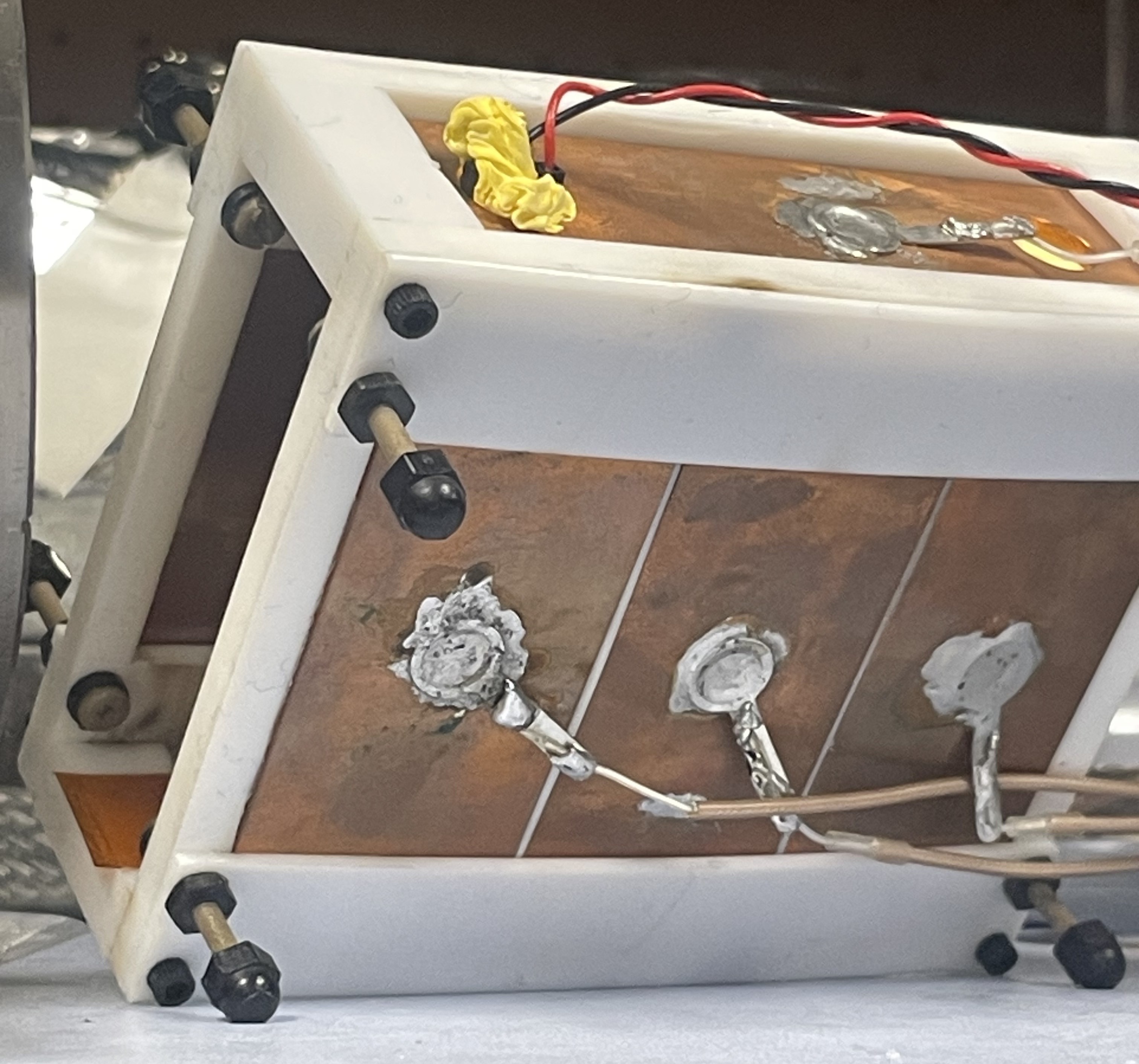}
    \caption{Wires supplying voltages are soldered directly to brackets on electrodes. The maximum voltage supplied is $\pm 6$ kV.}
    \label{fig:boxwire}
\end{figure}

Electrode voltage potentials are supplied by the CAEN Model DT8034 8-channel power supply, featuring a long term (1 week) voltage stability of $\pm 1.2$ V, as well as maximum voltage and current settings of $\pm 6$ kV and $1$ mA, respectively. Hence, the maximum potential barrier available to set on the bouncing electrodes is $-6$ kV, and the maximum potential difference between drift electrodes is $12$ kV $\left(\Rightarrow |{\bf E}| = (12 \: {\rm kV})/(0.03175 \: {\rm m}) \approx 378 \: {\rm kV}/{\rm m}\right)$. Current flows via a safe high voltage (SHV) connector directly to brackets soldered onto the electrodes as shown in Figure~\ref{fig:boxwire}.

\begin{figure}[!h]
\centering
    \includegraphics[width=0.95\linewidth]{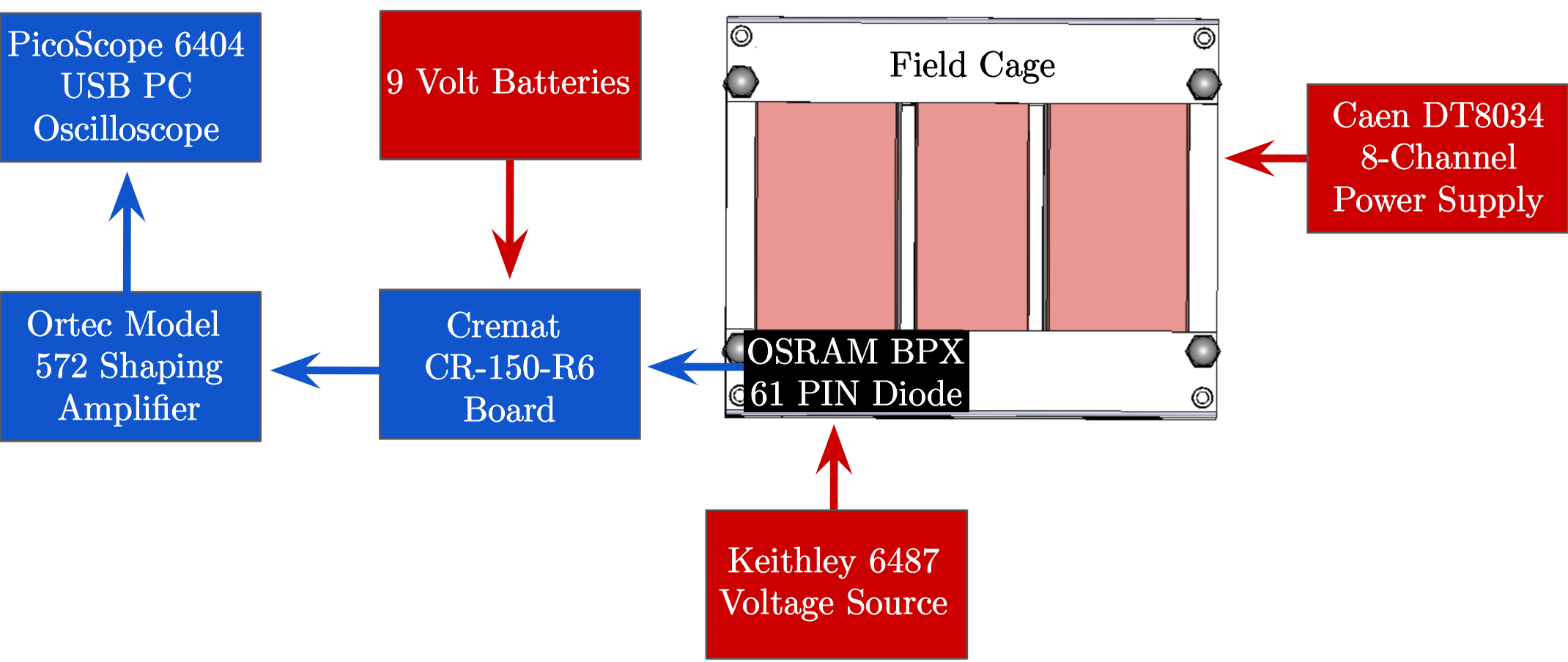}
    \caption{Block diagram showcasing various electrical components, including the detector (black), power supplies (red), and electronic readout chain (blue) for the setup.}
    \label{fig:block}
\end{figure}

 A reverse bias voltage of $25$ V is supplied to the BPX 61 diode by a Keithley 6487 picoammeter/voltage source. Operating at a reverse voltage slightly below the maximum setting (32 V) was done to prevent any undue damage to the modified detector. Pulses from the diode are read out using the commercially available Cremat CR-150-R6 Charge Sensitive Preamplifier (CSP) evaluation board. The board is configured following the company's recommendations for operating a device with dark current below 10 nA, and the CR-110-R2.2 CSP module was selected to attain the highest possible gain (1.4 ${\rm V}/{\rm pC}$) with low noise (equivalent noise charge of 200 electrons). Powering the board are two 9 V batteries so as to electrically isolate the device from any deviations/noise in wall power. Signals are then amplified at a second stage using an Ortec model 572 shaping amplifier with a gain of 25 and shaping time of 2 $\mu{\rm s}$. 
Finally, the output is read by a PicoScope 6404 USB PC oscilloscope; signal pulses exceeding a certain cutoff threshold are automatically saved by the device for further analysis. A block diagram of the electronic readout chain and power supplies is shown in Figure~\ref{fig:block}. In this study, the PicoScope waveform voltage range is selected to be $\pm 100$ mV with an offset of $-$80 mV. We set a trigger at $-$50 mV, meaning any waveform with an amplified pulse at least 30 mV above baseline is stored. 

\subsection{Layout and Magnet}\label{layoutandmagnet}
\begin{figure}[!h]
\centering
    \includegraphics[width=0.47\linewidth]{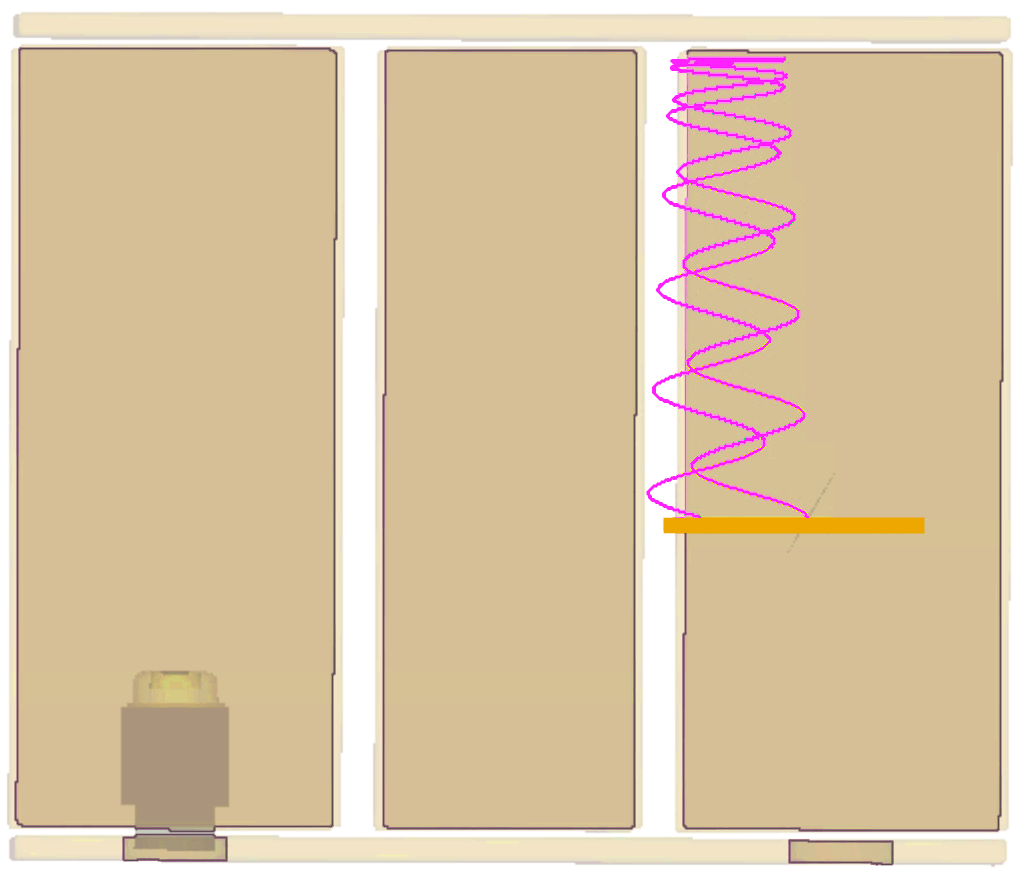}
    \caption{70 keV electron trajectory (magenta) terminating on the face of the source disk (yellow) when oriented parallel to the magnet pole faces. This example illustrates the low likelihood an emitted electron drifts far enough to clear the source disk radius (12.7 mm), motivating our decision to rotate the source by $60\degree$.}
    \label{fig:terminate}
\end{figure}

\begin{figure}[!h]
\centering
    \includegraphics[width=0.95\linewidth]{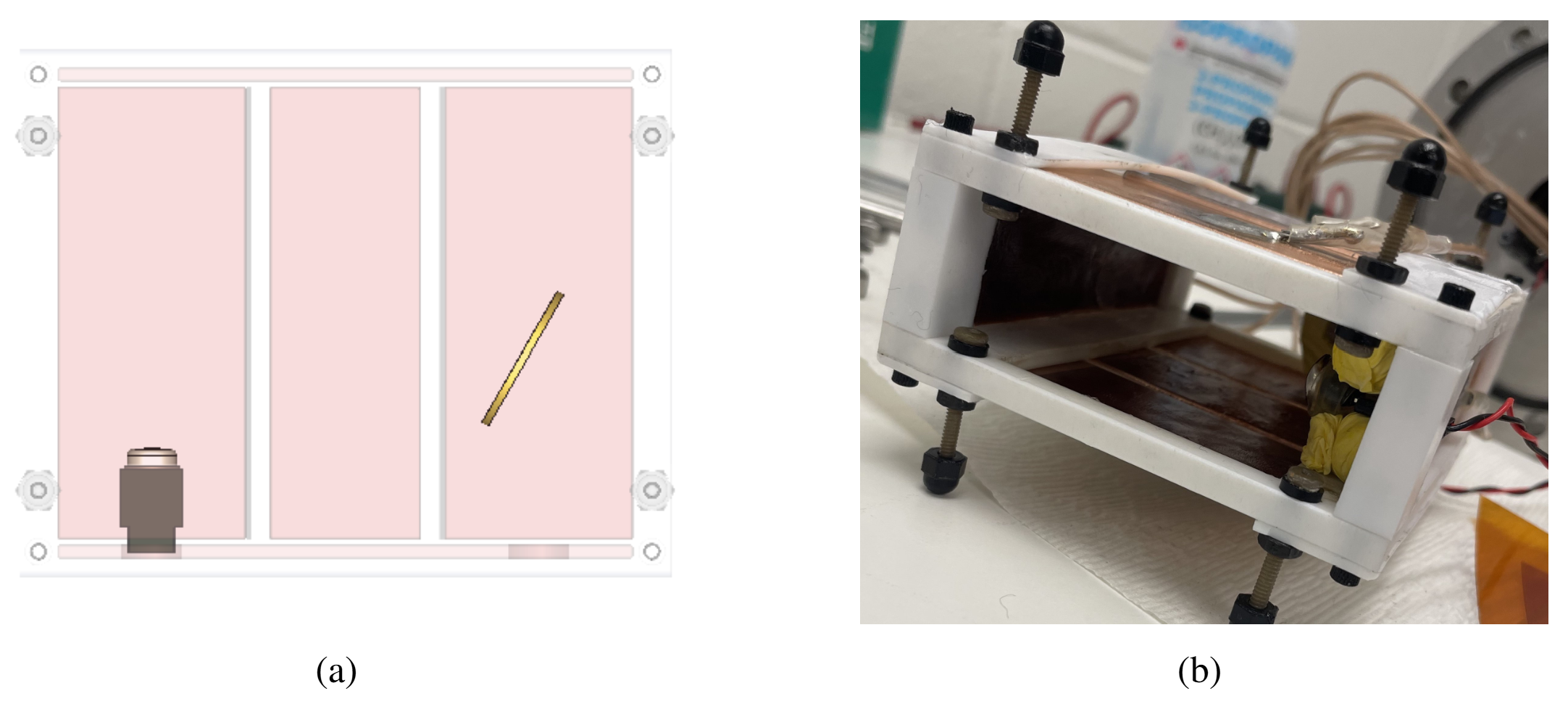}
    \caption{(a): Top-down field cage layout with $\ce{^{14}C}$ source disk (yellow) tilted at 60$\degree$ and diode (silver and black) fixed via injection hole. (b): Side profile of field cage layout, showcasing the diode's Teflon fixture and Kapton film (to mitigate scattering) lining electrode plates.}
    \label{fig:thecage}
\end{figure}

Using a Teflon block and tape, the $\ce{^{14}C}$ source disk is fixed between one of the outer pairs of drift electrodes and centered between the bouncing electrodes. The disk is tilted at 60$\degree$ with respect to the magnet pole faces. Considering drift velocity constraints imposed by the CAEN power supply, preliminary simulations indicated that nearly all emitted electrons terminated on the face of the source disk when oriented parallel to the magnet pole faces, meaning that electrons rarely drifted far enough upon the first bounce to clear the entire source disk radius $\left(12.7 {\: \rm mm} \right)$. An example of a terminated trajectory for a 70 keV electron is shown in Figure~\ref{fig:terminate}. Tilting the disk was the simplest and safest method to reduce the necessary drift length and improve clearance likelihood. Thus, electrons emitted from the 60$\degree$ tilted source only need to drift $6.35{\: \rm mm}$ to clear the disk. Repurposing the vestigial electron injection holes, the PIN diode detector is securely fixed between the other set of outer drift electrodes using Teflon tape. Inner electrode faces are additionally lined with Kapton polyimide film (76.2 $\mu$m thick) to mitigate electron scattering. Materials with a higher atomic number have a greater probability of scattering incident particles, making Kapton ($Z \approx 7$) a suitable cover for copper ($Z = 29$) to reduce scattering frequency. Top-down and side profiles of the field cage layout are shown in Figures~\ref{fig:thecage}.

\begin{figure}[!h]
\centering
    \includegraphics[width=0.5\linewidth]{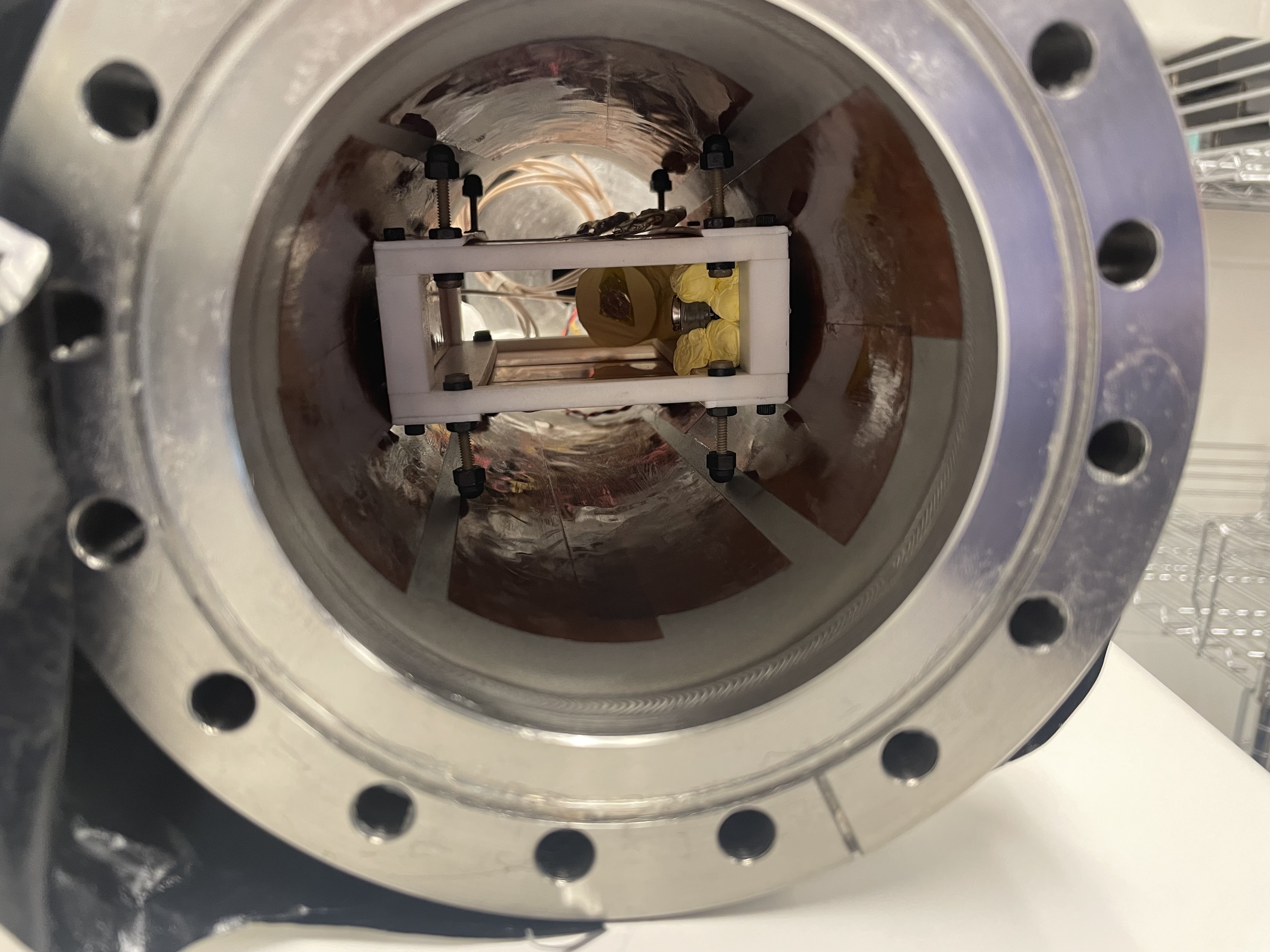}
    \caption{Field cage situated in $10.2 {\: \rm cm}$ OD stainless steel vacuum chamber with inner walls lined by Kapton polyimide film to prevent electrical shorting between the chamber and electrode plates.}
    \label{fig:chamberinside}
\end{figure}

\begin{figure}[!h]
\centering
    \includegraphics[width=0.5\linewidth]{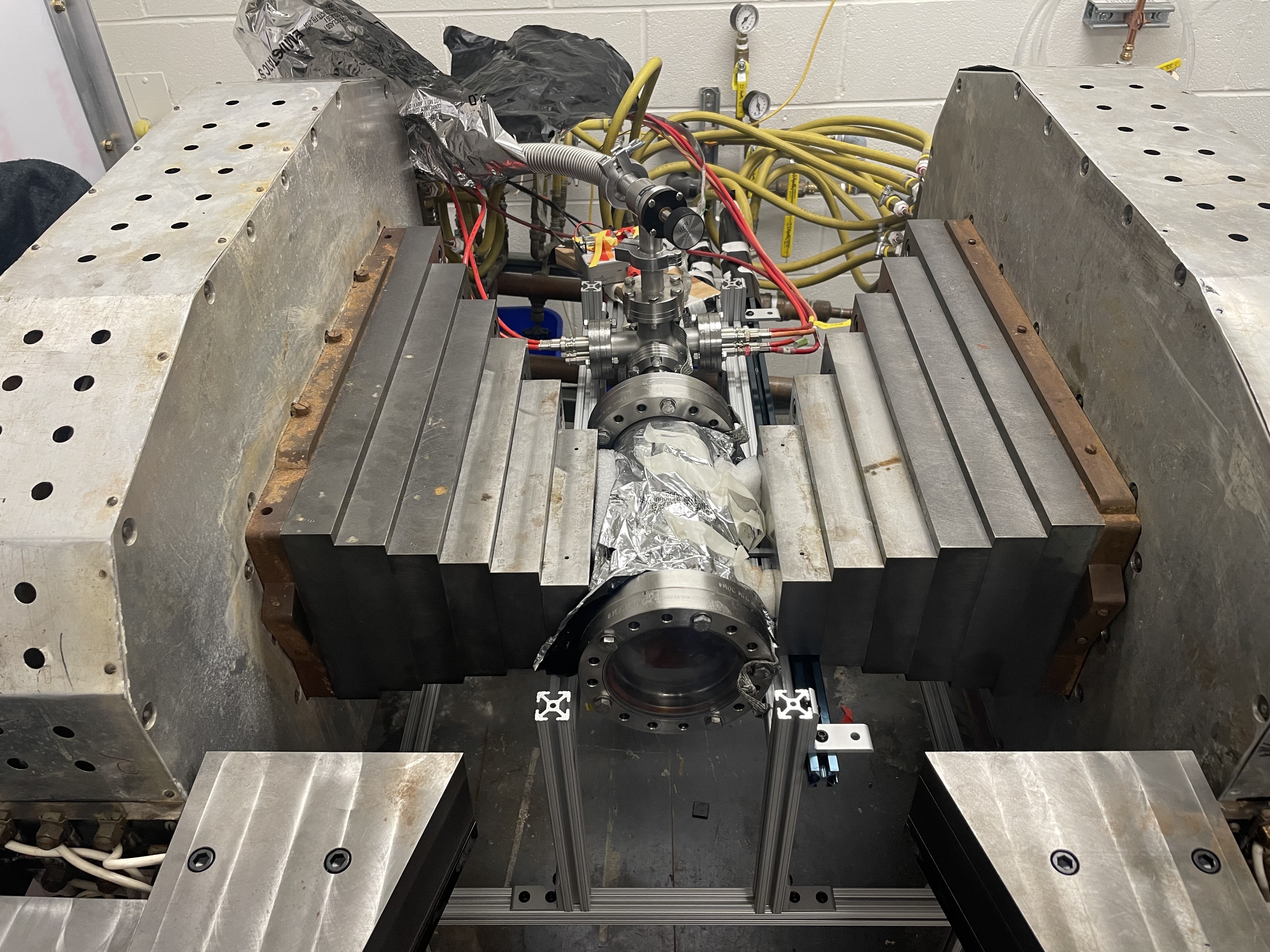}
    \caption{Field cage and vacuum chamber fixed between magnet pole faces with auxiliary components (connectors for voltage, electronic readout, and vacuum pumping) placed away from the high magnetic field region (top of image). The pressure is maintained at roughly 0.1 mPa, and the glass window flange is covered with blackout sheets when in operation. In this study, a uniform magnetic field of 160 mT is produced between the pole faces.}
    \label{fig:chamberandmag}
\end{figure}

The field cage is placed within a $10.2 {\: \rm cm}$ OD stainless steel vacuum chamber. The inner wall is lined with Kapton film (adhered by vacuum grease) to prevent electrical shorting between the chamber and electrode plates, as shown in Figure~\ref{fig:chamberinside}. The bulk of the chamber is placed adjacent to the magnet's pole faces (rotationally fixed by foam blocks), with connectors for vacuum pumping, voltage supply, and readout situated farther away from the high magnetic field region. Using a molecular drag pump, the system is held at a pressure of roughly $0.1$ mPa. A picture of the full system layout is shown in Figure~\ref{fig:chamberandmag}. The glass window flange allows for precise alignment of the field cage in relation to the magnet. When the system is operating, the flange is covered with blackout sheets.

The magnet consists of six coils (three on each side) powered by Walker Scientific Inc. HS-96-302 and HS-1575-4SS power supplies. Under maximum current conditions, a uniform magnetic field of $\approx 1$ T is generated between the pole faces. For this study, we operate at a much lower setting to produce a field of $\approx 160$ mT. Given that ${\bf v}_{E} = E/B$ for orthogonal fields, a lower magnetic field increases the maximal drift velocity and hence likelihood for electrons to clear aforementioned obstacles. Under these conditions, maximal drift velocity is given as $E/B \approx 2.4 \times 10^6 $  m/s. Assuming the distance between the detector and source disk to be roughly 60 mm and ignoring effects due to the cyclotron radius, the estimated electron time of flight when operating all six drift electrodes at $\pm 6$ kV is $(0.06 \: {\rm m})/(2.4\times 10^6 \: {\rm m}/{\rm s})= 25$ ns.

\section{Procedure}
\subsection{Operational Settings}

\begin{figure}[!h]
\centering
    \includegraphics[width=0.90\linewidth]{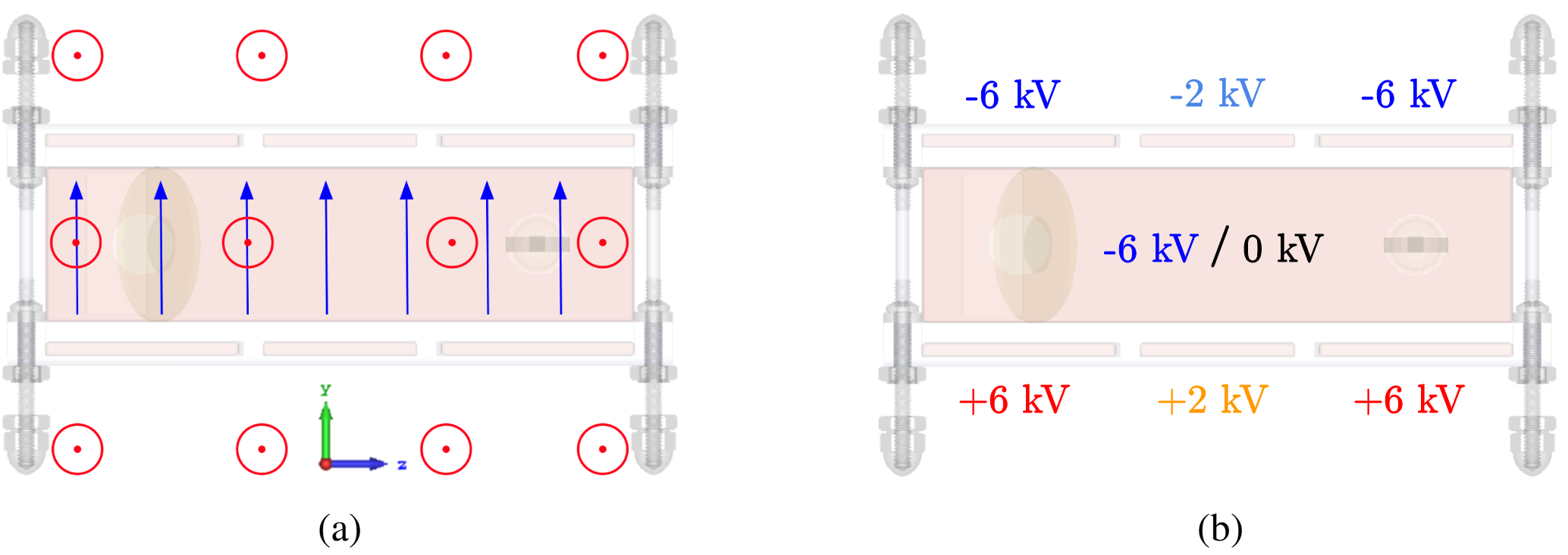}
    \caption{(a): Orientation of ${\bf B}$ (red) and ${\bf E}$ (blue) to generate ${\bf E}$ $\times$ ${\bf B}$ drift towards detector. ${\bf E}$ is variably set by voltage differences between electrodes. (b): Voltage settings for electrode plates to obtain $1/3$ maximal drift in the center region. Data is recorded operating the bouncing electrodes at both $-$6 and 0 kV; the difference between these settings represents the event rate of successfully transported electrons.}
    \label{fig:diagram1}
\end{figure}

Following the orientation of the magnetic field (pointing leftward in Figure~\ref{fig:chamberandmag}), an \emph{upward} electric field is required to produce ${\bf E}$ $\times$ ${\bf B}$ drift directed toward the diode. This implies the lower and upper outer drift electrodes are respectively set to $+$6 and $-$6 kV for maximal drift velocity and source disk clearance. The center pair of drift electrodes is set at a smaller voltage difference to achieve slower drift (for instance $+$2 and $-$2 kV for $1/3$ maximal drift velocity). Field orientation and an example voltage setting for the field cage are shown in Figure~\ref{fig:diagram1}. Data is recorded operating the bouncing electrodes at \emph{both} $-$6 and 0 kV. As electrons are capable of traveling through the system (and being detected by the PIN diode) by means of scattering off of the bouncing electrode plates, the difference in rates between these settings represents the event rate of successfully transported (i.e., bouncing) electrons. For any voltage setting, data-taking for the 0 kV bouncing electrode mode is performed between repeated runs of the $-$6 kV mode to ensure consistent behavior of the system.

\subsection{Waveform Analysis}

\begin{figure}[!h]
\centering
    \includegraphics[width=0.95\linewidth]{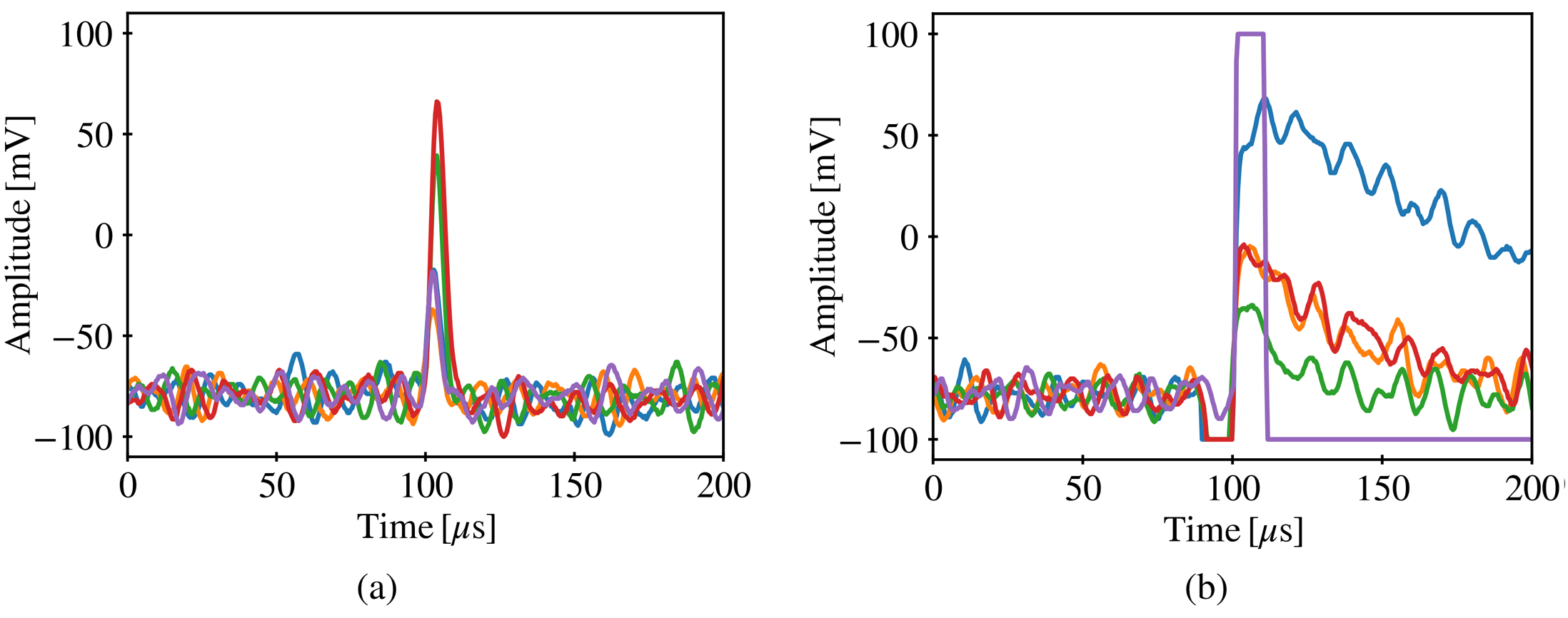}
    \caption{Example waveforms (a) passing and (b) not passing the mean and standard deviation selection criteria outlined in Table~\ref{tab:crit}. As mentioned previously, the baseline of waveform output is offset by -80 mV. The peak height above baseline of passing waveforms is proportional to the energy deposition of the impinging particle on the detector.}
    \label{fig:passornopass}
\end{figure}

Waveform information stored by the PicoScope for triggered pulses is used to filter spurious events (including noise chirps or inverted pulses due to unstable electrical ground) from the data. We apply lenient ($\geq$ 98\% efficient for most data runs) selection criteria to multiple regimes of the 200 $\mu$s time domain saved for each event (100 $\mu$s before and after the pulse trigger). These criteria applied to the mean ($\mu$) and standard deviation ($\sigma$) of the waveform during various selected time domains are outlined in Table~\ref{tab:crit}. {\color{black}The general motivation for the requirements is that triggered noise characteristically has either displaced $\mu$ directly preceding or following the trigger at 100 $\mu$s, or has extremely large/small $\sigma$ in the later time steps. These specific ranges and values were determined from fine-tuning the denoising of multiple waveform subsets.} Examples of waveforms that pass and do not pass such requirements are shown in Figures~\ref{fig:passornopass}. For passing waveforms, the amplitude of the pulse height is proportional to the energy deposition of the impinging particle.

\setlength{\arrayrulewidth}{0.4pt} 
\setlength{\tabcolsep}{10pt}      
\renewcommand{\arraystretch}{1.5} 

\begin{table}[h!]
\caption{Selection criteria on waveform mean ($\mu$) and standard deviation ($\sigma$) over selected time periods of $200\: \mu$s saved output for each triggered event. In most data runs, $\geq$ 98\% of events pass these selection criteria.}
\centering
\vspace{5pt}
\begin{tabular}{|c|c|c|}
    \hline  
    {\bf Time Domain} & {\bf Selection Criterion} \\ \hline
    $[90\:\mu {\rm s}, 96 \:\mu {\rm s}]$ & $\mu > -87 \: {\rm mV}$ \\ 
    $[116\:\mu {\rm s}, 124 \:\mu {\rm s}]$ & $\mu \leq -77 \: {\rm mV}$  \\ 
    $[140\:\mu {\rm s}, 180\: \mu {\rm s}]$  & $2 \: {\rm mV} < \sigma \leq 80 \: {\rm mV}$ \\ 
    \hline   
\end{tabular}
\label{tab:crit}
\end{table}

\begin{figure}[!h]
\centering
    \includegraphics[width=0.5\linewidth]{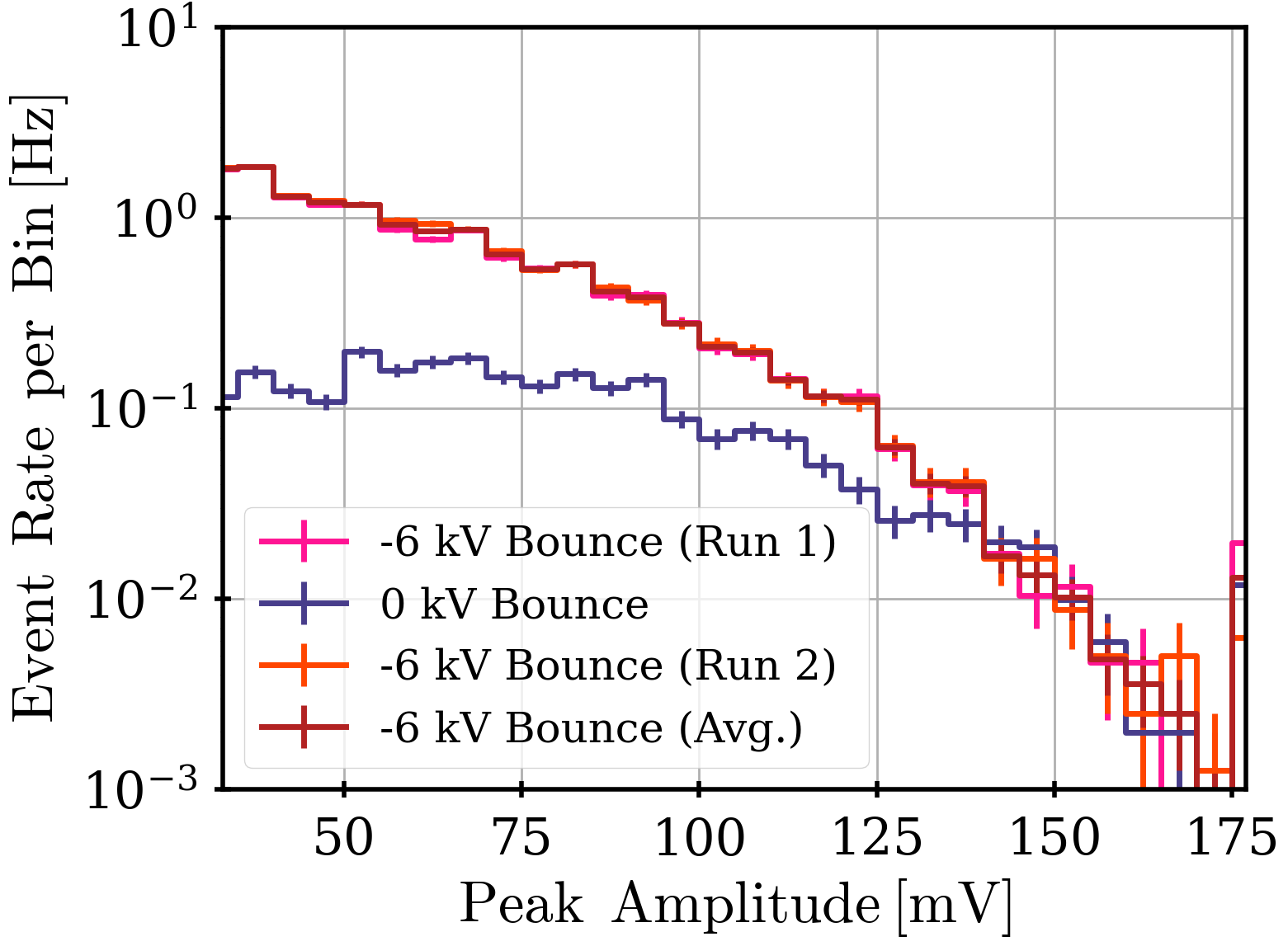}
    \caption{Binned event rates for $\pm$2 kV central drift electrode setting for various bouncing electrode settings. Event rates are obtained by binning waveforms according to peak height and dividing each bin value by the data acquisition time (normally $\approx$ 900 seconds). The difference in event rate (particularly in smaller bins) between $-6$ kV and $0$ kV bounce electrode settings indicates successful electron transport. Multiple $-6$ kV bounce runs are taken to ensure consistent system behavior. Uncertainties on each bin are assigned according to Poisson statistics.}
    \label{fig:2kv_evtrates}
\end{figure}

Once waveform analysis is performed, events are binned into a histogram according to their peak height (with the offset baseline shifted back to 0 mV). These bin counts are then divided by the data acquisition time (normally on the order of 900 seconds) to obtain an event rate for various slow drift voltage settings. An example of such outputs including Poisson statistic error bars is shown in Figure~\ref{fig:2kv_evtrates} for the central drift electrode setting of $\pm$2 kV. The significantly higher --- and consistent --- event rate with bouncing plates operated at $-$6 kV compared to 0 kV indicates successful transport of electrons. Statistical separation is most pronounced in the smaller voltage bins, while the spectra tend to converge at larger voltage bins --- suggesting that the detection of higher-energy electrons is more heavily influenced by scattering. Events with pulse heights exceeding the range of the PicoScope are included in the highest voltage bin (known as an overflow bin).

\subsection{Modes of Operation}\label{modesss}
\begin{figure}[!h]
\centering
    \includegraphics[width=.9\linewidth]{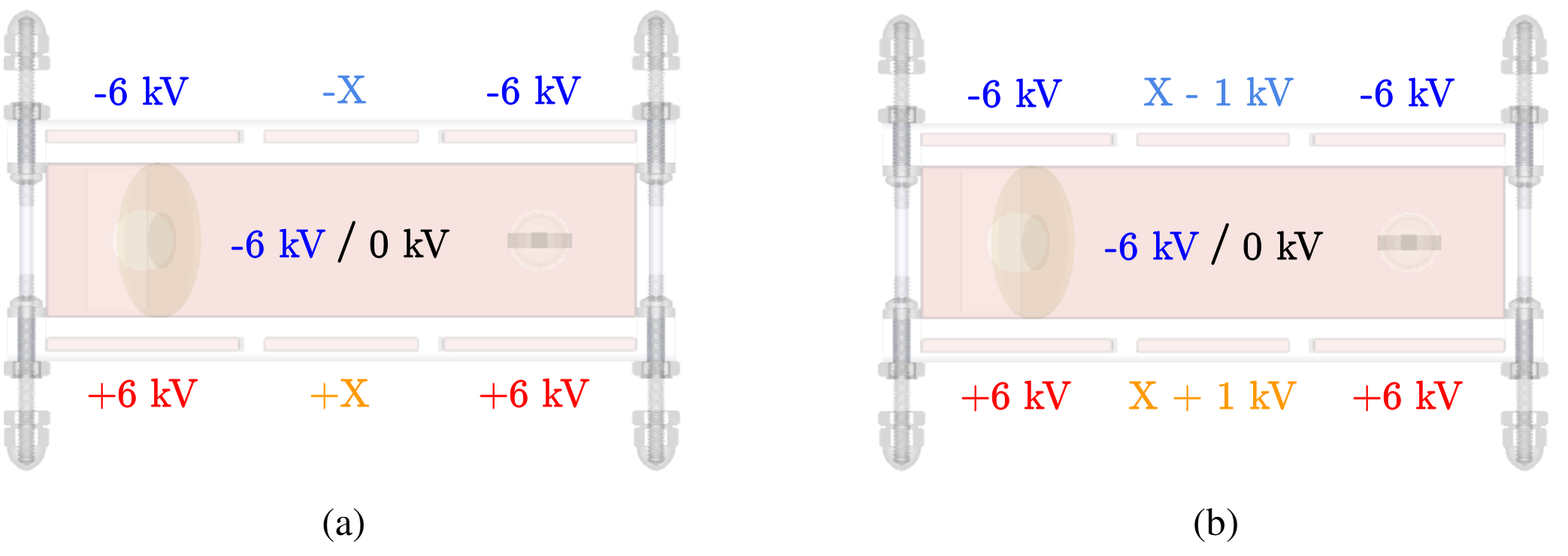}
    \caption{(a): Symmetric slow drift setting, in which upper and lower central electrode plates are set to the same voltage ${\bf X}$ though opposite polarities, and are gradually decreased. (b): Window voltage scan setting, in which potential difference between upper and lower central electrode plates is maintained at a fixed value (e.g., 2 kV), though offset by various {\it midplane} voltage ${\bf X}$. These are the two data-taking methods used to probe characteristics of the demonstrator.}
    \label{fig:symmetricandscan}
\end{figure}

\begin{figure}[!h]
\centering
    \includegraphics[width=1\linewidth]{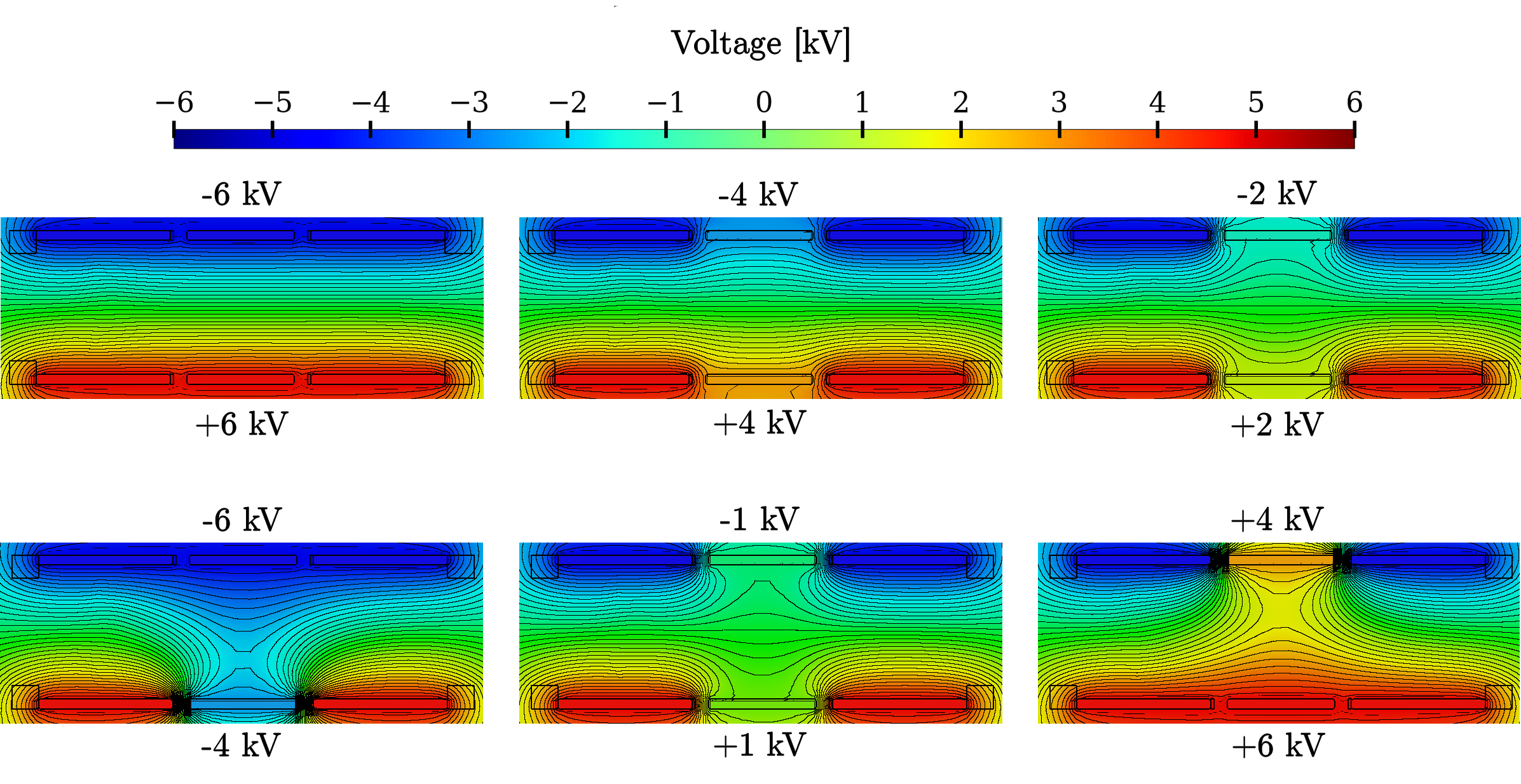}
    \caption{Potential lines for various symmetric slow drift (top row) and 2 kV window voltage scan (bottom row) settings. For symmetric slow drift, potential lines disperse outward as the center potential difference decreases, meaning lower event rates are expected for smaller potential differences. In the case of the voltage scans, only certain potential lines remain continuous; a maximal event rate is expected at a negative midplane voltage due to the angled source disk.}
    \label{fig:sympot}
\end{figure}

{\color{black}In order to create multiple, coherent datasets to compare with simulation, we record event rates with the ${\bf E}$ $\times$ ${\bf B}$ slow drift demonstrator according to two operational modes:} \emph{symmetric slow drift} and \emph{voltage scan} (both shown in Figure~\ref{fig:symmetricandscan}). {\color{black}With direct application to PTOLEMY,} symmetric slow drift studies the change in event rate as the magnitude of the central drift plate voltages is {\it symmetrically} reduced (keeping a potential of 0 V at the midplane). {\color{black}This corresponds to a slowed drift speed in the central region.} Using CST,  the top row of Figure~\ref{fig:sympot} shows that potential lines disperse outward as the center potential difference decreases (displayed is a cross-section of the center plane in the demonstrator, though is representative of the field in general). With fewer potential lines remaining continuous, lower total event rates are expected for smaller central drift settings (${\bf E}$~$\times$~${\bf B}$ does no work, necessitating the guiding center of an emitted electron to follow potential lines of the orthogonal ${\bf E}$ field). {\color{black}Symmetric slow drift data on its own, however, would not make an airtight case for system function. An observed reduction in event rates could simply be a result of a lower voltage (less noisy) operational environment, but a correspondence with simulation would nonetheless be seen.}

Alternatively, a voltage scan maintains a constant potential difference in the central plates (for instance a $2$ kV `window') but alters the midplane potential (central plate settings of $+$1 and $-$1 kV, $+$2 and 0 kV, $+$3 and $+$1 kV, etc.). As seen in the bottom row of Figure~\ref{fig:sympot}, only certain potential lines remain continuous depending on the midplane potential. {\color{black}A scan of this nature with a stronger magnetic field could be used to spatially profile the activity of a radioactive source, with resolution defined according to window size.} For our setup, the angled source disk --- which more frequently emits particles pointed \emph{towards} the diode than \emph{away} --- transmits electrons more often with guiding centers in the upper half of the field cage. A maximal event rate is thus expected at a negative midplane voltage. Results are presented for both 1 kV and 2 kV window voltage scans.

\section{Results}\label{results}
\subsection{Symmetric Slow Drift}\label{symres}

\begin{figure}[!h]
\centering
    \includegraphics[width=0.48\linewidth]{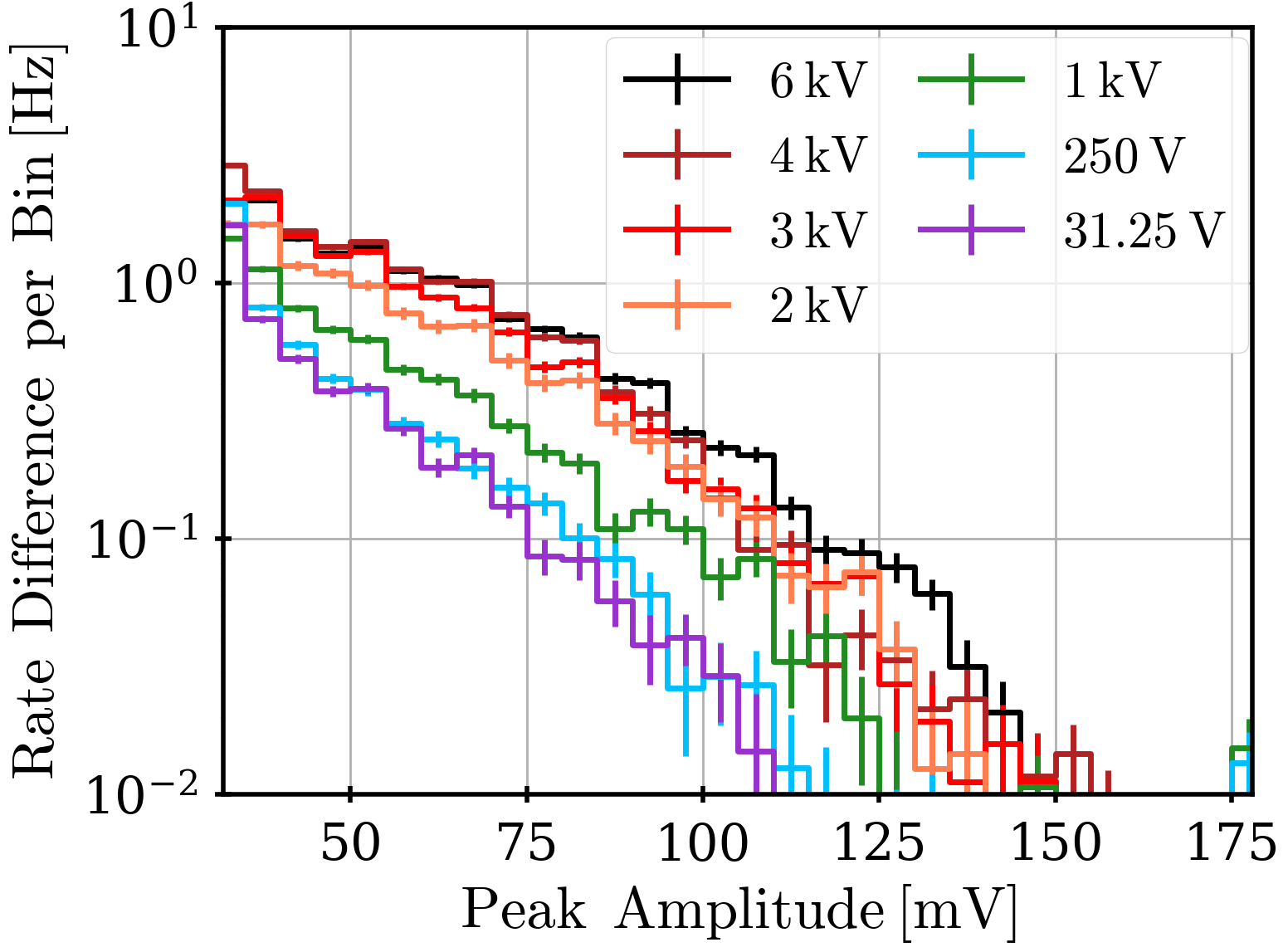}
    \caption{Binned event rates of successfully transported (i.e., difference in event rate of $-6$ kV and $0$ kV bounce settings) electrons for various symmetric slow drift settings. Settings are reported as the potential magnitude set to both central drift plates. Those with a larger potential have a higher event rate for nearly every voltage bin and longer high voltage tail.}
    \label{fig:slowdrift}
\end{figure}

\begin{figure}[!h]
\centering
    \includegraphics[width=0.48\linewidth]{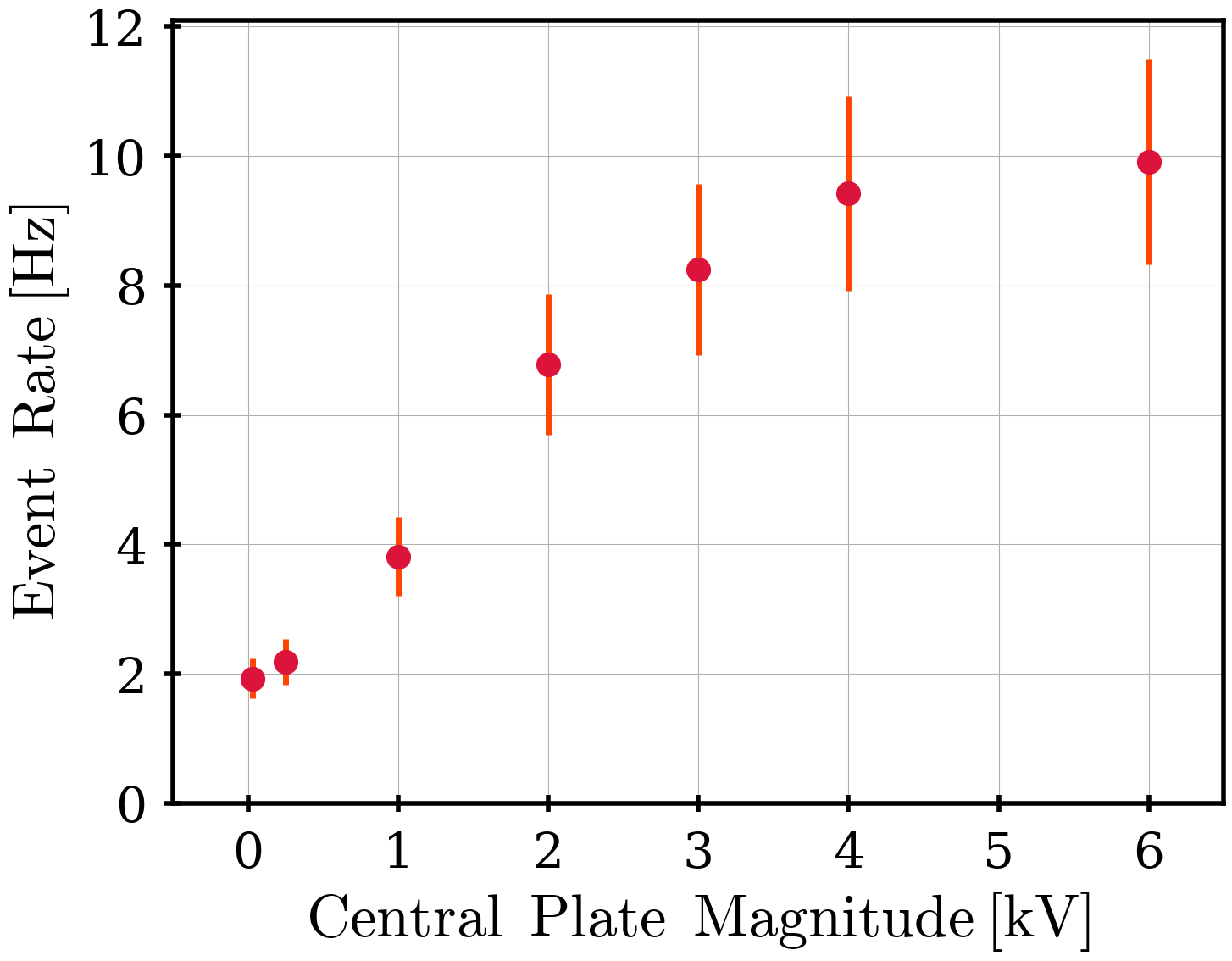}
    \caption{Cumulative event rates (above 45 mV) of successfully transported electrons for various symmetric slow drift settings. The distribution features regions of steep and plateauing event rate increase, a maximal event rate of 10 events per second at $\pm 6$ kV, and a nonzero event rate at $\pm 0$ kV. A fractional uncertainty of 16\% is assigned to each event rate based on repeated measurements.}
    \label{fig:slowdriftcumulative}
\end{figure}

Event rates, including error bars from Poisson statistics, of successfully transported electrons are shown in Figure~\ref{fig:slowdrift} for various symmetric central drift electrode voltage settings. Settings are reported according to the voltage magnitude set to both central drift plates (i.e., 6 kV corresponds to lower and upper plates set at $\pm 6$ kV, etc.). Central electrode settings with a larger potential difference have a higher event rate for nearly every voltage bin and longer high voltage tail. To study the total event rate for each potential setting, we sum values in all bins above 45 mV (a threshold well above the noise level of the system and consistent with the PIN diode calibration using the beta spectrometer) in Figure~\ref{fig:slowdriftcumulative}. {\color{black} Intuitively, a smaller center plate voltage magnitude corresponds to a reduction in event rate.} The distribution additionally exhibits regimes of steep and plateauing event rate as a function of the center potential difference (and a maximal event rate of roughly 10 events per second at $\pm 6$ kV). Notably, potential differences approaching zero maintain a \emph{nonzero} event rate, suggesting the residual field produced by the outer/bounce plates facilitates drift for small central plate settings. {\color{black}These less intuitive features are due to the geometry of the system and are mirrored in simulation as seen in Section~\ref{datacomp}.}


\subsection{Window Voltage Scans}
\begin{figure}[!h]
\centering
    \includegraphics[width=1\linewidth]{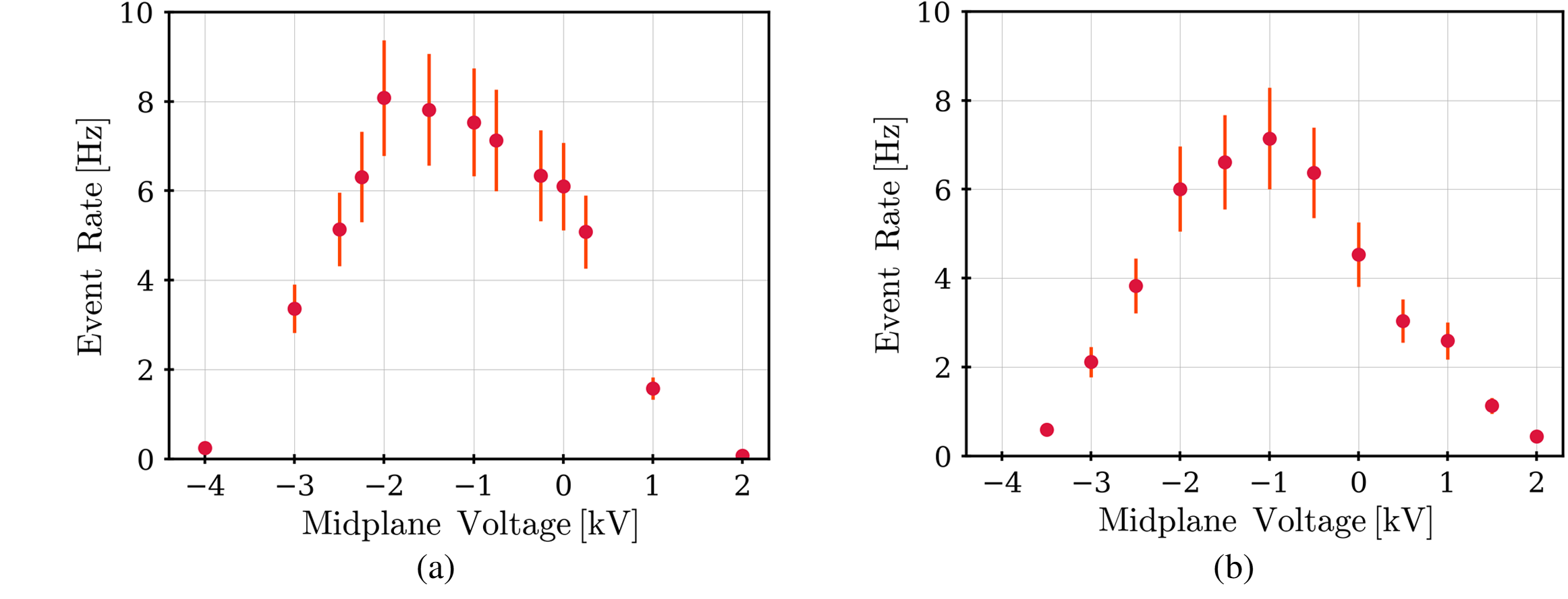}
    \caption{Cumulative event rates (above 45 mV) of successfully transported electrons for various (a) 2 kV and (b) 1 kV window settings plotted as a function of midplane voltage. Both distributions are roughly centered around $-1$ kV, and the 1 kV window event rates are smaller than the 2 kV window counterpart for the same midplane voltage setting. Again, a fractional uncertainty of 16\% is assigned to each event rate based on repeated measurements.}
    \label{fig:scan2kand1k}
\end{figure}

Similarly, cumulative event rates are tabulated and plotted as a function of the midplane voltage setting (i.e., average of top and bottom center plates) for both $2$ kV and $1$ kV window scans in Figure~\ref{fig:scan2kand1k}. Both distributions are roughly centered around $-$1 kV and event rates of the 1 kV scan are chiefly smaller than their 2 kV scan counterparts. {\color{black}As with Section~\ref{symres}, attributes of these distributions are largely influenced by the geometry of the system (particularly source disk angling). }In Section~\ref{datacomp}, we compare these results with CST counterparts to verify the accuracy of the simulated model.

\subsection{Uncertainty}
{\color{black} We establish a fractional uncertainty of 16\% on data points by repeating measurements of nearly all settings. The value corresponds to the minimum fractional width on each point in the original dataset such that at least $66\%$ of the total repeated measurements fall within the respective interval (i.e., assuming the distribution is roughly Gaussian, an interval containing $\approx 66\%$ of events corresponds to 1 $\sigma$). This broadly encompasses systematic uncertainties associated with slight variability in chamber alignment, variations in noise level, electric/magnetic field stability, and deviations in detector response (due to self-heating among other phenomena). It, however, does {\it not} account for all systematic uncertainties for the physical system, in particular ignoring source alignment, detector alignment, trajectory impediment from Teflon structures, magnetic field uniformity, changes in detector response for various electromagnetic fields, active region size of the detector, and other parameters which do not change between measurements. We consider these uncertainties small enough to ignore, though emphasize the PTOLEMY detector will be rigorously characterized in this regard. Uncertainty on the relative height between the source and detector is addressed in simulation results.}

\section{Particle Simulations}\label{simulations}

\subsection{Particle Spectra}
\begin{figure}[!h]
\centering
    \includegraphics[width=0.65\linewidth]{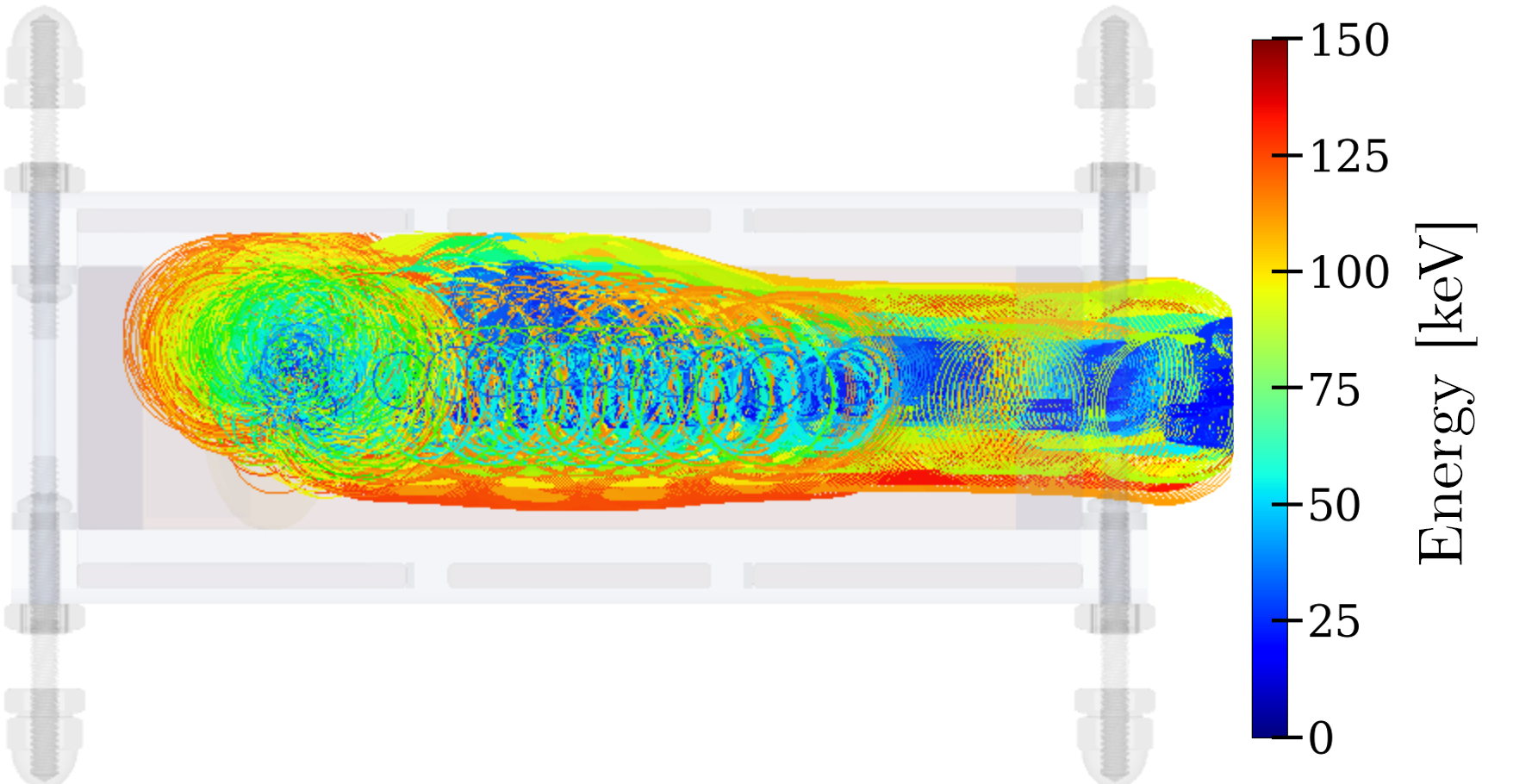}
    \caption{Asymmetry of emitted electron trajectories under symmetric $+$2 and $-$2 kV central voltage settings. This serves as a qualitative confirmation in simulation of our assessment that angling the source disk favors electrons with guiding centers in the upper half of the field cage. The computational intensity of trajectory information limits more quantitative assessments.}
    \label{fig:asymtraj}
\end{figure}

\begin{figure}[!h]
\centering
    \includegraphics[width=0.47\linewidth]{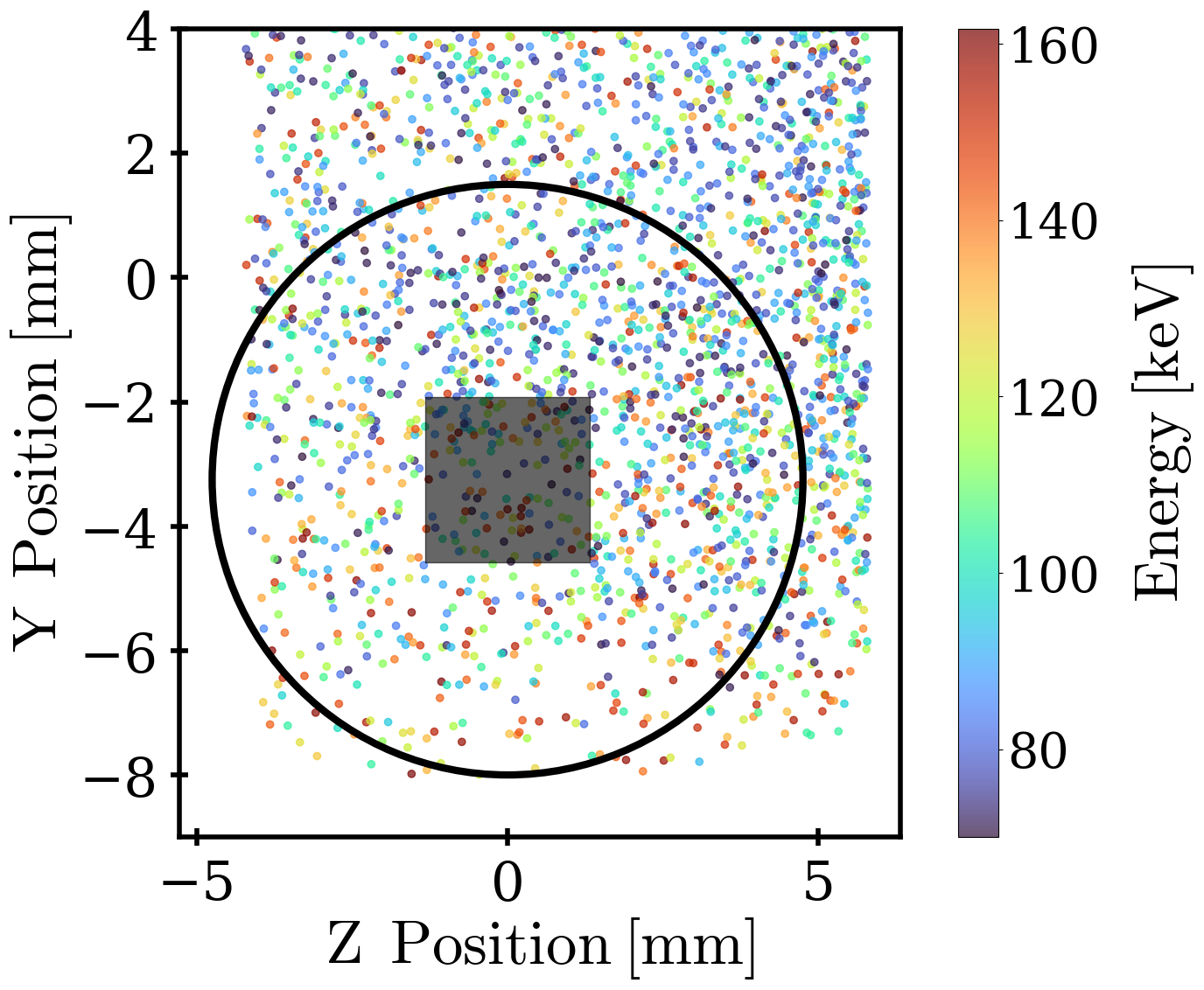}
    \caption{Final location and energy of emitted electrons using 2D particle monitor output in CST. This feature allows for 1 million electrons to be simulated under each setting. The Y position corresponds to the relative height with respect to the electron emission location. The $-3.25$ mm offset of the diode (overlaid in black) was chosen as its resulting spectrum most closely resembles data, and the relative height is mildly ambiguous in the physical setup. For visibility, only electrons with energy above $70$ keV are shown.}
    \label{fig:energy_map}
\end{figure}

Importing design files into the software, the demonstrator was replicated in CST, and simulations were run under a multitude of voltage settings. To qualitatively understand the phase space of transmitted electrons, we record trajectories of emitted particles. This, for instance, allows us to confirm the proposed asymmetrical transmitted phase space of electrons as suggested in Section~\ref{modesss} (Figure~\ref{fig:asymtraj}). Trajectory information, however, is highly computationally intensive and places stringent limits on the number of electrons we can simulate. For more quantitative estimates, we instead record the final location of each particle using the 2D particle monitor feature. This enables us to simulate 1 million electrons (2$\pi$ solid angle and uniform energy ranging between 0 and 160 keV), corresponding to a few seconds of emission from the $\ce{^{14}C}$ source. From the monitor, we save electrons whose terminal location corresponds to the active region of the diode ($\leq$ 0.04\% emitted). The output provides details on the phase space of electron transmission, including incident angle with the diode surface, which was the motivation for angling the diode for energy calibration as specified in Section~\ref{PINDIODE}. An example of such an output with the diode face overlaid is shown in Figure~\ref{fig:energy_map}. The active area of the diode is situated 3.25 mm downward relative to the height of the $\ce{^{14}C}$ emission source. As the relative height between the two in the physical setup is mildly ambiguous (i.e., the radioactive material on the source disk could have been deposited anywhere within a few mm from the disk center), this offset was chosen as its resultant spectrum most closely resembles data.

\begin{figure}[!h]
\centering
    \includegraphics[width=0.5\linewidth]{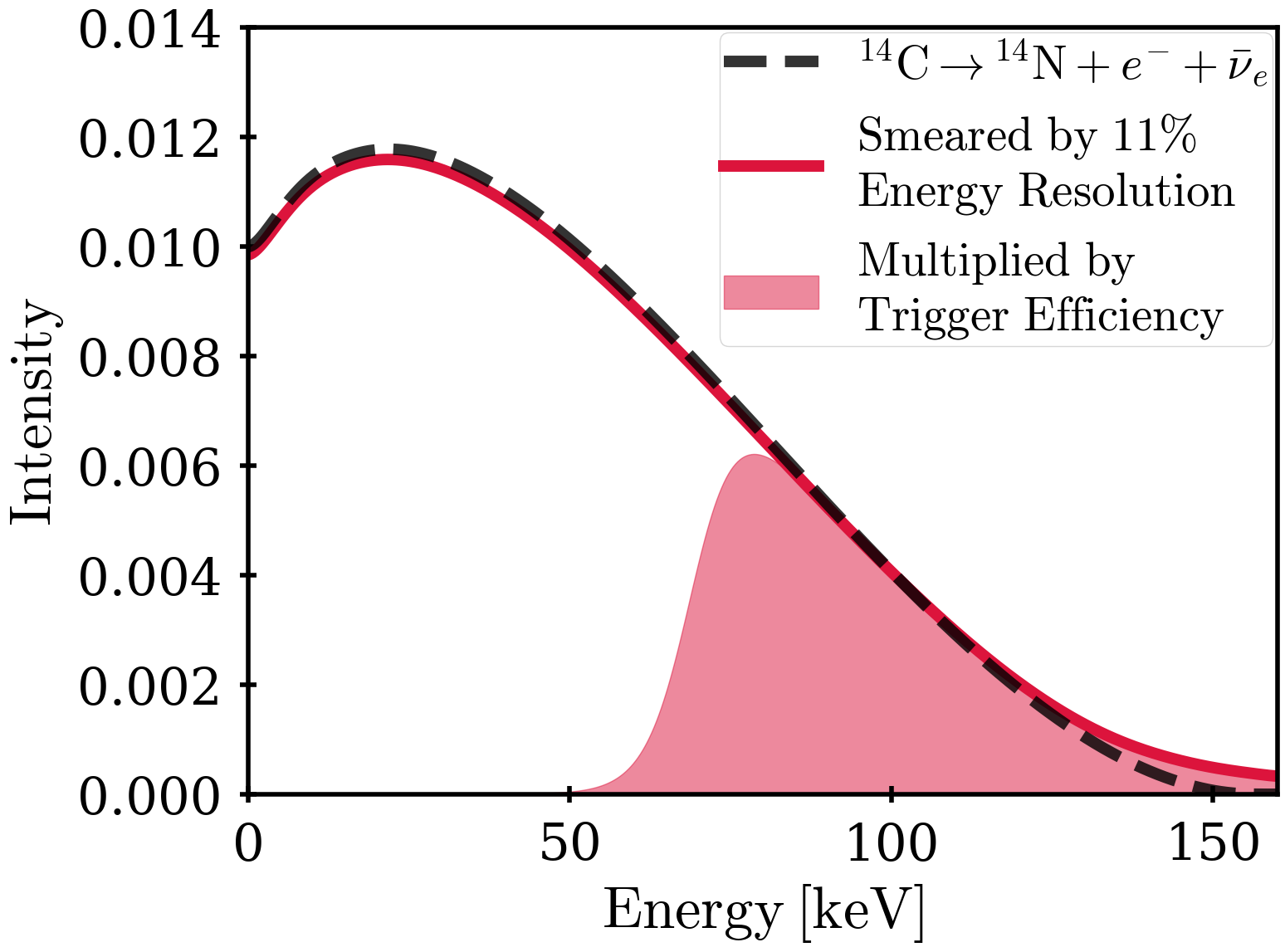}
    \caption{Product of PIN diode trigger efficiency curve and $\ce{^{14}C}$ $\beta$ spectrum (smeared by 11\% energy resolution of detector). This curve is discretized and used as a weighting to scale the uniformly-emitted simulation results such that they can be compared directly to data.}
    \label{fig:c14_smear}
\end{figure}

To obtain accurate event rates for comparison with data, counts are first binned by energy. These bin values are then divided by the total number of emitted electrons (1 million) and multiplied by the source activity (5 $\mu$Ci). This, however, is the event rate as given from a \emph{uniform} energy emission source. To obtain an event rate reflective of the source emission spectrum, each bin is finally weighted according to the normalized, discretized product of the $\ce{^{14}C}$ $\beta$ spectrum (smeared by the detector's 11\% energy resolution) and the PIN diode trigger efficiency curve shown in Figure~\ref{fig:c14_smear}. The sum of these bins produces event rates which we compare directly to data.

\subsection{Comparison to Data}\label{datacomp}
\begin{figure}[!h]
\centering
    \includegraphics[width=1\linewidth]{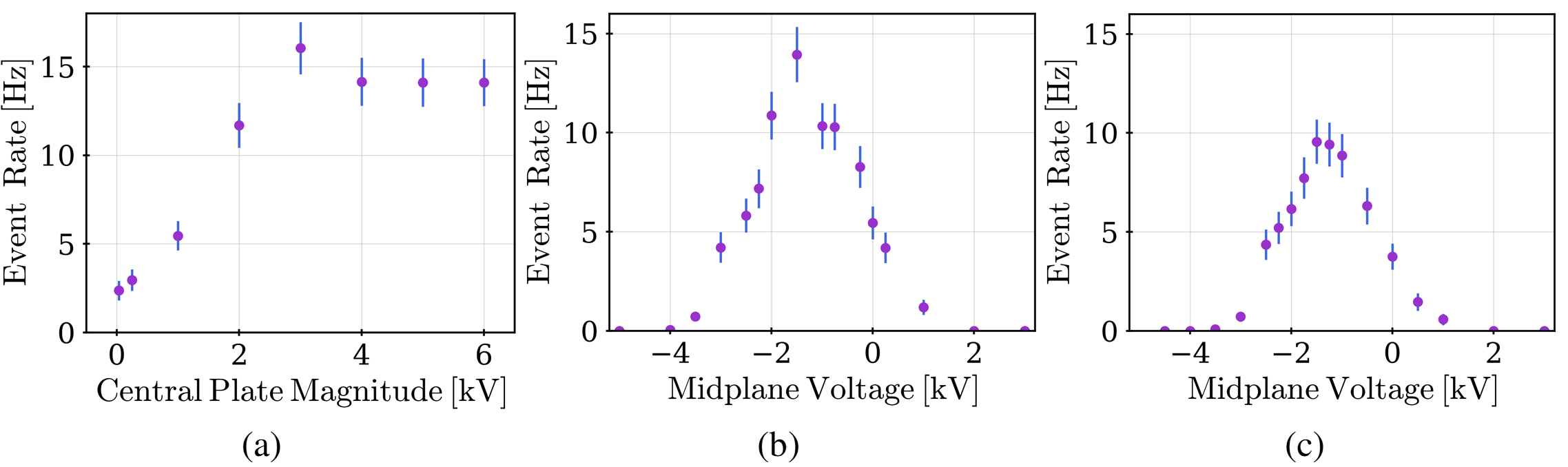}
    \caption{CST simulated event rates scaled according to Figure~\ref{fig:c14_smear} for (a) symmetric, (b) 2 kV window, and (c) 1 kV window settings. Uncertainty is assigned according to Poisson statistics. Event rates obtained for each setting are higher than in data.}
    \label{fig:symstraight}
\end{figure}

\begin{figure}[!h]
\centering
    \includegraphics[width=1\linewidth]{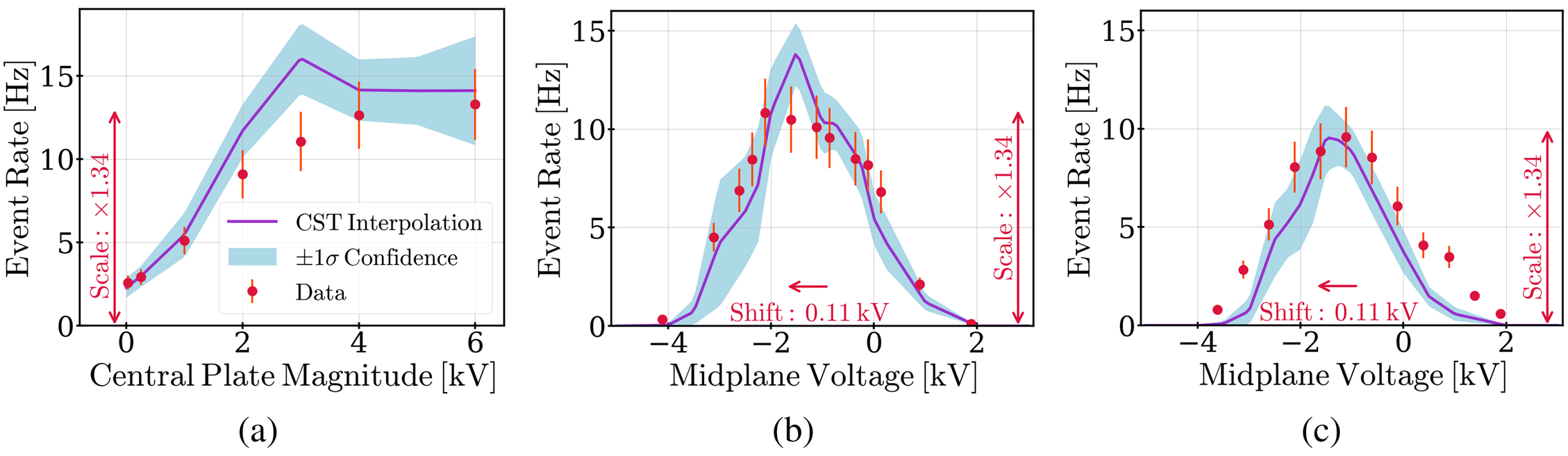}
    \caption{Combined least-squares regression fit of data (red) to interpolated CST rates (purple) for (a) symmetric, (b) 2 kV window, and (c) 1 kV window settings. Scaling factor to data is 1.34 and leftward horizontal offset (applied to 1 kV and 2 kV window results) is 0.11 kV. A $\pm 1$ $\sigma$ confidence interval for simulated event rates accounts for Poisson statistical fluctuations and adjusting the diode active region location by $\pm 0.5$ mm. A more detailed systematics study would need to be undertaken in order to fit according to an uncertainty-based metric such as likelihood. Nonetheless, a correspondence between the simulated and detected event rates by a small scaling factor demonstrates the validity of the simulated model.}
    \label{fig:fits}
\end{figure}

Cumulative event rates obtained from CST are presented in Figure~\ref{fig:symstraight},
exhibiting slightly higher values than their Section~\ref{results} counterparts. We demonstrate a correspondence between the two trios of distributions in Figure~\ref{fig:fits}, fitting according to least squares for a scaling factor and horizontal offset (for the 1 kV and 2 kV windows). Event rates from CST are interpolated, and a $\pm$1 $\sigma$ confidence interval is tabulated accounting for Poisson statistical fluctuation and deviations in event rates from adjusting the acceptance window by $\pm 0.5$ mm. A more rigorous study of systematic uncertainties would be required to fit according to likelihood or another uncertainty-based metric. The horizontal offset of 0.11 kV is an indication that the $\ce{^{14}C}$ source disk is not perfectly centered with respect to the drift electrodes, which may partially explain the lower event rates in data compared to simulation for small symmetric settings in Figure~\ref{fig:fits}. The scaling factor of 1.34 on the detected event rate serves as a general approximation to demonstrate the validity of the simulated model. The $\approx$ 25\% discrepancy in event rate is likely due to a confluence of reasons including system alignment, impedance from Teflon structures, smaller-than-expected diode active region, etc. In the following section, we proceed with time of flight estimation.


\subsection{Time of Flight Estimation}\label{sec:tof}

\begin{figure}[!h]
\centering
    \includegraphics[width=1\linewidth]{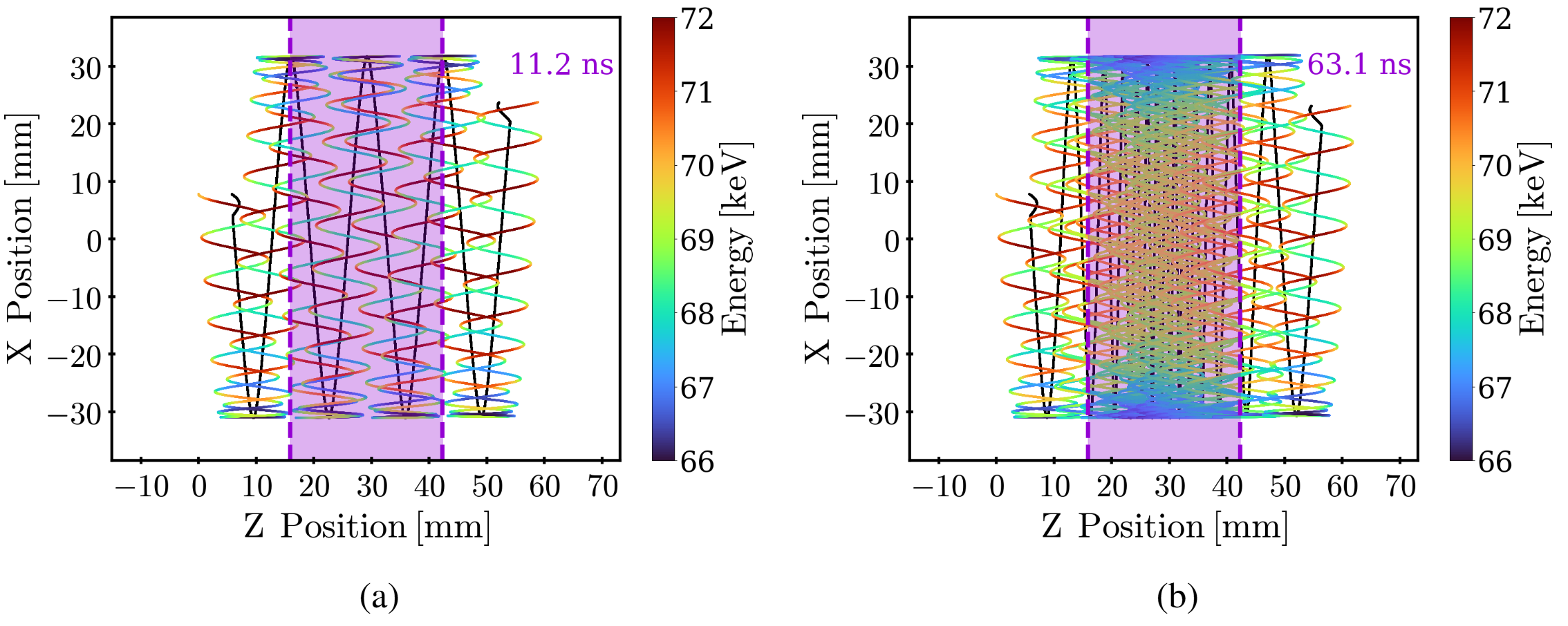}
    \caption{Simulation of 70 keV electron transported with center plates set to (a) $\pm 6$ kV and (b) $\pm 31.25$ V.  Guiding center of trajectory is plotted in black, and the slow drift region is shaded in purple. The amount of time the guiding center is situated in the slow drift region is printed in the top-right corner. The slow drift setting allows one to trap an electron roughly 5-6 times longer than the maximally fast drift setting ($63.1$ ns in comparison to $11.2$ ns).}
    \label{fastandslowtraj}
\end{figure}

\begin{figure}[!h]
\centering
    \includegraphics[width=1\linewidth]{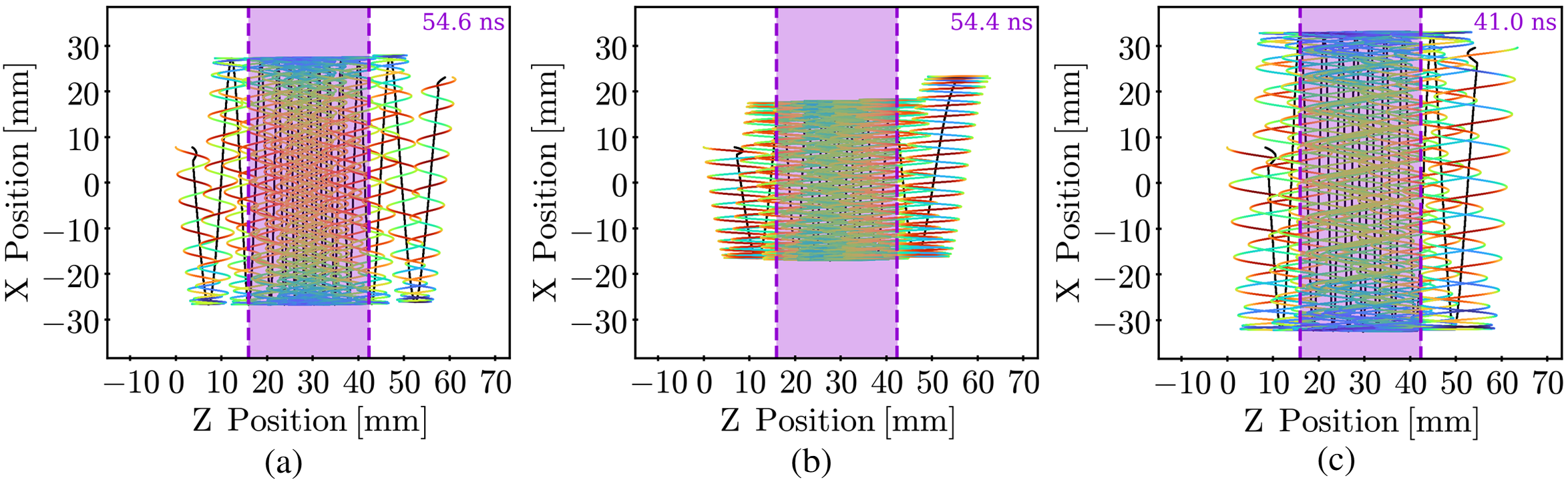}
    \caption{Guiding center and trajectory under $\pm 31.25$ V center plate conditions for simulated electrons with energy (a) 30 keV, (b) 100 keV, and (c) 150 keV. The respective times of flight in the center plate region are 54.6 ns, 54.4 ns, and 41.0 ns. In general, time of flight is roughly 54-66 ns for the majority of the spectra, though decreases slightly for exceedingly energetic ($\geq 130$ keV) electrons.}
    \label{fig:moretof}
\end{figure}

To evaluate the time of flight of particles in the slow drift region of the demonstrator, we export trajectory information from CST under the studied $\pm6$ kV and $\pm31.25$ V central drift settings. {\color{black}As the largest and (roughly) smallest attainable central magnitudes, these trajectories correspond to the fastest and slowest possible drift speeds.} The guiding center at each point is tabulated by averaging the electron location over one cyclotron period before/after the point of interest.\footnote{In the relativistic regime, the period in which a particle completes one full orbit due to cyclotron motion is given as $T = \frac{2\pi \gamma m}{q B}$, where $m$ is mass, $\gamma$ is the Lorentz factor, $q$ is charge, and $B$ is magnetic field strength.} Time of flight in the slow drift region is then determined based on how long the guiding center is situated between the $25.4{\: \rm mm}$ central plates. Figure~\ref{fastandslowtraj} shows that a 70 keV electron can be {\color{black}maximally} slowed to less than $1/5$ of the fastest drift velocity, being trapped for over 60 ns in the center plate. {\color{black}One would naively expect a much slower drift speed for such a proportionally small voltage setting; unfortunately, for small center plate voltage magnitudes, the fringe electric field produced by other plates dominates the central region, yielding a fairly high lower bound on drift speed.} The observed drift speed in the trajectory is consistent with $E/B$ as evaluated from the simulated fields at each point. Deviations in time of flight occur due to the amount of time a given trajectory spends near the bouncing plates (those that bounce more and reflect nearer to the bouncing plate experience a weaker central drift field on average than those that bounce fewer times and farther away, therefore having a slightly longer drift time). Simulations with various particle energies confirm the slow drift time of flight largely falls within the range of 54-66 ns, only decreasing slightly in the case of $\geq$ 130 keV electrons (at which point the cyclotron radius is so large that the electron spends a considerable amount of time under the influence of the outer plate potentials even when its guiding center is situated near the middle). Additional slow drift trajectories are shown in Figure~\ref{fig:moretof}. As the electric field produced in the center region is heavily impacted by potentials set on the outer plates for this setup, increasing the dimensions of the slow drift region, thereby reducing fringe field effects, would allow for a slow drift time of flight on the order of $\mu $s and largely reduce variations for a given trajectory.

\section{Outlook}
The control of $\beta$-decay ${\bf E}$ $\times$ ${\bf B}$ electron drift {\color{black}in a defined volume} has been exhibited. Under current operating conditions, we can slow drift speed by a factor of 5 (containing electrons for over 60 ns), though would theoretically be able to increase this by orders of magnitude by altering the dimensions of the field cage design. {\color{black}Manipulation of drift speed to trap electrons as such is an essential feature for PTOLEMY to suppress backgrounds and perform a sensitive differential energy measurement of the tritium endpoint to determine $m_\beta$. The method of voltage scanning as presented additionally holds value as a technique to spatially profile the activity of a radioactive source.} Serving as a demonstration for the RF region, future extensions of this project will look to incorporate electron injection from a low magnetic field region and instrument an antenna array to perform CRES measurements akin to the studies performed by the Project 8 Collaboration~\cite{AshtariEsfahani_2024} and under development for PTOLEMY~\cite{iwasaki2024cresbasednondestructiveelectronmomentum,Pesce}.

\appendix
\acknowledgments
 This material is based upon work supported by the National Science Foundation Graduate Research Fellowship Program under Grant No.DGE-2039656. Any opinions, findings, and conclusions or recommendations expressed in this material are those of the author(s) and do not necessarily reflect the views of the National Science Foundation. A. Colijn and J. Mead are supported by the Dutch Research Council (NWA.1292.19.231). I. Rago is supported by the MIUR
program (CUP:B81I18001170001). C. Tully is supported by the John Templeton Foundation (\textnumero$\,62313$). The
authors acknowledge the support of the INFN CSN-V.

\bibliographystyle{JHEP}
\bibliography{biblio.bib}
\input{supplementary}

\end{document}

%% file: supplementary.tex
\newpage
\section*{Supplementary Material}\label{Supplementary}

\section{Beta Spectrometer}

\begin{figure}[!htbp]
\begin{center}
\includegraphics[width=3.6in]{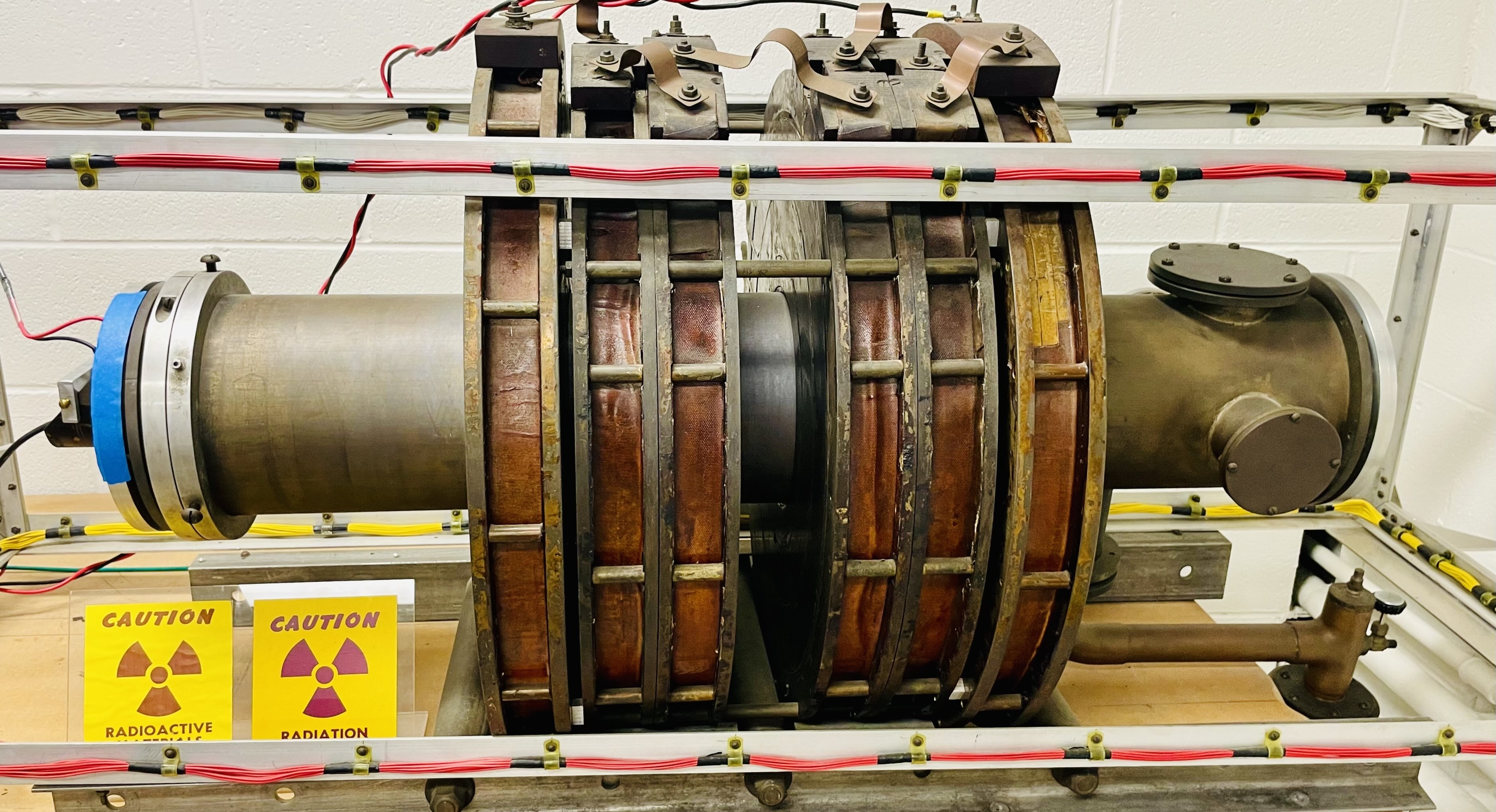}
\caption{\small
Photograph of the full spectrometer setup used in the experiment. The system includes a vacuum chamber, magnetic lenses, and shielding components arranged to guide and focus electron trajectories for precise beta spectroscopy.
\label{fig:spectrometer} }
\end{center}
\end{figure}
\begin{figure}[!htbp]
\begin{center}
\includegraphics[width=4.5in]{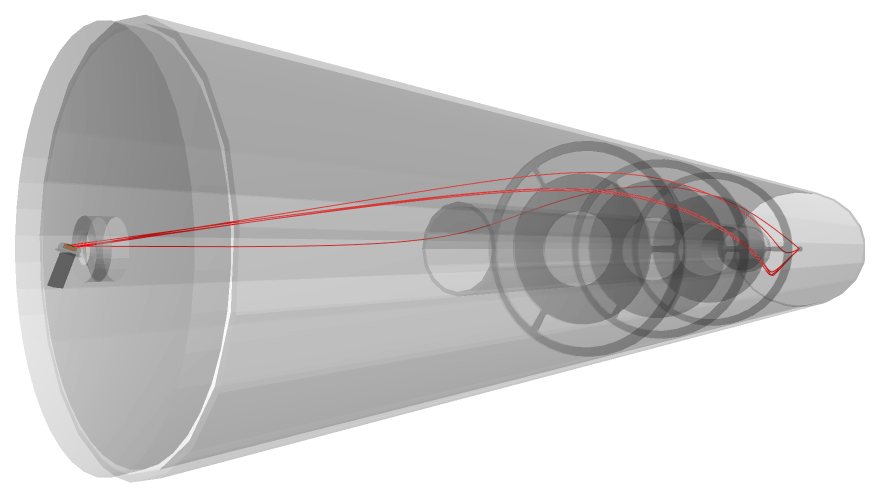}
\caption{\small
Simulated electron trajectories within the spectrometer. The red lines represent 70~keV electrons traversing the magnetic lenses, influenced by applied magnetic fields with 3~A coil currents.
\label{fig:spectrometerCST} }
\end{center}
\end{figure}

\vspace{-0.23in}  
\noindent The beta spectrometer used in this study is a magnetic lens spectrometer, designed to analyze beta decay energy spectra by selectively filtering electrons based on their momenta~(Figure~\ref{fig:spectrometer}). It consists of a vacuum chamber, a magnetic lens system, and the PIN diode as the detector. The vacuum chamber ensures minimal scattering of beta particles at about 1~mTorr. The spectrometer’s magnetic lens system, configured in a long solenoidal arrangement, generates a cylindrically symmetric magnetic field, guiding beta particles along helical trajectories. By adjusting the field strength, electrons of specific momenta are focused onto the detector, as illustrated in the simulated trajectories in Figure~\ref{fig:spectrometerCST}. This configuration sets a constant momentum resolution of 2.45\%~($\sigma/\mu$).
\vspace{-.2in}  

\begin{figure}[!htbp]
\begin{center}
\includegraphics[width=2.in]{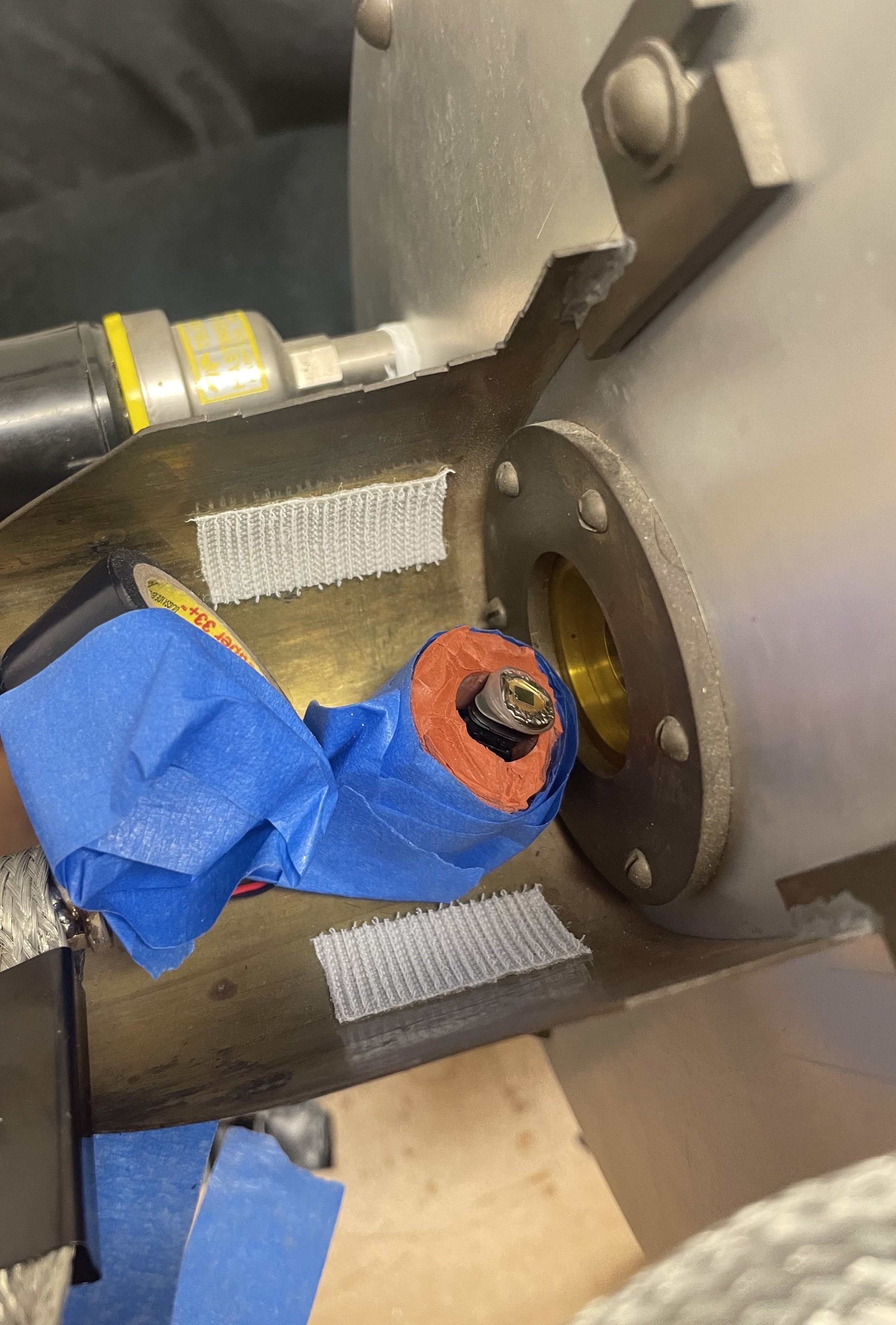}
\includegraphics[width=2.6in]{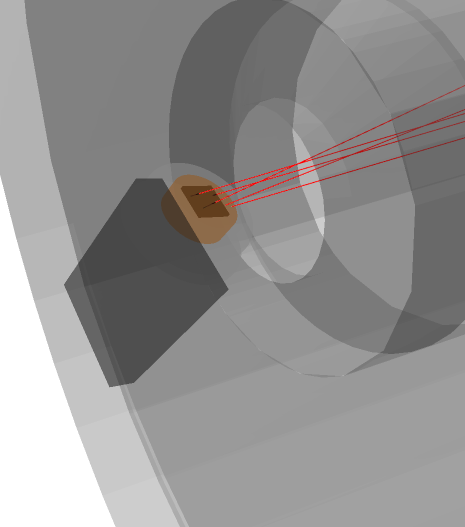}
\caption{\small
(Left) Calibration setup showing the PIN diode and magnetic lens system during alignment and testing. (Right) Simulated electron trajectories under calibration conditions, focusing on the PIN diode region where electron paths converge and interact with the detector’s surface for precise momentum calibration.
\label{fig:spectrometerPIN} }
\end{center}
\end{figure}

The PIN diode was positioned at the spectrometer’s exit plane to record beta particles after magnetic field selection, as shown in Figure~\ref{fig:spectrometerPIN} (left). The system is calibrated using a $^{137}$Cs source, which provides a well-defined reference energy for momentum calibration. A Kapton thin-film window at the spectrometer’s entry minimizes electron energy loss while maintaining vacuum integrity, ensuring accurate energy calibration. Additionally, external magnetic trim coils compensate for stray fields, ensuring a stable and reproducible spectrometric response.

